\documentclass[11pt,preprint]{aastex}

\usepackage{placeins}

\shorttitle{Protostellar Accretion}
\shortauthors{Zhu \& Hartmann}

\begin{document}

\title{Non-steady Accretion in Protostars}

\author{Zhaohuan Zhu\altaffilmark{1}, Lee Hartmann\altaffilmark{1}, and
Charles Gammie
\altaffilmark{2,3}}

\altaffiltext{1}{Dept. of Astronomy, University of Michigan,
500 Church St., Ann Arbor, MI 48105}
\altaffiltext{2}{Dept. of Astronomy, University of Illinois
Urbana-Champaign,
1002 W. Green St., Urbana, IL 61801}
\altaffiltext{3}{Dept. of Physics, University of Illinois
Urbana-Champaign}

\email{zhuzh@umich.edu, lhartm@umich.edu, gammie@illinois.edu}

\newcommand\msun{\rm M_{\odot}}
\newcommand\lsun{\rm L_{\odot}}
\newcommand\msunyr{\rm M_{\odot}\,yr^{-1}}
\newcommand\be{\begin{equation}}
\newcommand\en{\end{equation}}
\newcommand\cm{\rm cm}
\newcommand\kms{\rm{\, km \, s^{-1}}}
\newcommand\K{\rm K}
\newcommand\etal{{\rm et al}.\ }
\newcommand\sd{\partial}
\newcommand\mdot{\rm \dot{M}}
\newcommand\rsun{\rm R_{\odot}}

\begin{abstract}

Observations indicate that mass accretion rates onto low-mass
protostars are generally lower than the rates of infall to their
disks; this suggests that much of the protostellar mass must be
accreted during rare, short outbursts of rapid accretion.  We
explore when protostellar disk accretion is likely to be highly
variable.  While constant $\alpha$ disks can in principle adjust
their accretion rates to match infall rates, protostellar disks are
unlikely to have constant $\alpha$.  In particular we show that
neither models with angular momentum transport due solely to the
magnetorotational instability (MRI) nor gravitational instability
(GI)  are likely to transport disk mass at protostellar infall rates
over the large range of radii needed to move infalling envelope
material down to the central protostar.  We show that the MRI and GI
are likely to combine to produce outbursts of rapid accretion
starting at a few AU. Our analysis is consistent with the
time-dependent models of Armitage, Livio, \& Pringle (2001) and
agrees with our observational study of the outbursting object FU
Ori.

\end{abstract}

\keywords{accretion disks, stars: formation, stars: pre-main sequence}

\section{Introduction}

The standard model of low-mass star formation posits the free-fall
collapse of a protostellar molecular cloud core to a protostar plus
disk during times of a few times $10^{5}$~yr (e.g., Shu, Adams, \&
Lizano 1987), consistent with the statistics of protostellar objects
in Taurus (Kenyon \etal 1990, 1994). To build up a star over these
timescales requires a time-averaged infall rate of order $2 \times
10^{-6} - 10^{-5} \msunyr$, rates typically used in calculations of
protostellar properties at the end of accretion (Stahler 1988;
Hartmann, Cassen, \& Kenyon 1997). The numerical simulations of
dynamic star cluster formation by \cite{2003MNRAS.339..577B} found
that stars and brown dwarfs formed in burst lasting $\sim$
2$\times$10$^{4}$ years, implying infall rates of $\sim$ 10$^{-4}$
to 10$^{-5} \msunyr$. However, the accretion luminosity implied by
such infall rates is considerably higher than typical observed
protostellar luminosities (Kenyon \etal 1990, 1994).  This
``luminosity problem'' can be solved temporarily by piling up
infalling matter in the circumstellar disk.  Most of the mass must
eventually be accreted onto the star, however.  This requires major
accretion events that are sufficiently short-lived that protostars
are usually observed in quiescence.

This picture of highly time-dependent accretion is supported by
observations. Individual knots in jets and Herbig-Haro objects, thought
to be the result of outflows driven by accretion energy, argue for
substantial disk variability (e.g., Bally, Reipurth, \& Davis 2007). The
FU Ori objects provide direct evidence for short episodes of rapid
accretion in early stages of stellar evolution, with accretion rates of
$10^{-4} \msunyr$ or more (Herbig 1977; Hartmann \& Kenyon 1996), vastly
larger than typical infall rates of $\lesssim 10^{-5} \msunyr$ for
low-mass objects (e.g., Kenyon, Calvet, \& Hartmann 1993; Furlan \etal
2008).

The mechanism driving FU Ori outbursts is not yet clear.  A variety of
models have been proposed: thermal instability (TI; Lin \& Papaloizou
1985; Bell \& Lin 1994); gravitational instability (GI; Vorobyov \& Basu
2005, 2005); gravitational instability and activation of the
magnetorotational instability (MRI; Armitage, Livio, \& Pringle 2001;
also Gammie 1999 and Book \& Hartmann 2005); and even models in which
planets act as a dam limiting downstream accretion onto the star (Clarke
\& Syer 1996; Lodato \& Clarke 2004).  Our recent analysis based on
Spitzer IRS data (Zhu \etal 2007) led us to conclude that a pure TI
model cannot work for FU Ori.

In view of the complexity of the problem and the physical uncertainties
we adopt a schematic approach.  We start with the (optimistic)
assumption that protostellar accretion can be steady.  We then show that
the GI is likely to dominate in the outer disk, while the MRI is likely
to be important in the inner disk, and that mismatches between the GI
and MRI result in non-steady accretion for expected protostellar infall
rates.  \footnote{GI and MRI here means turbulent states initiated by
gravitational or magnetic instabilities, respectively.}  Our analysis
agrees with the results found in the time-dependent outburst model of
Armitage \etal (2001), and is consistent with our empirical analysis of
the outbursting system FU Ori (Zhu \etal 2007).  Although our results
depend on simplified treatments of the GI and MRI, the overall picture
is insensitive to parameter choices.  We predict that above a critical
infall rate protostellar disk accretion can be (relatively) steady;
observational confirmation would help constrain mass transport rates by
the GI and the MRI.

\section{Overview}

A disk with viscosity \footnote{We use ``viscosity'' as shorthand
for internal, localized transport of angular momentum by turbulence.
We will also make the nontrivial assumption that external torques
(e.g. MHD winds) can be neglected.} $\nu$ will evolve at cylindrical
radius $r$ on a timescale \be t_{\nu} \sim r^2/\nu \en If this is
comparable to the timescales over which mass is being added to the
disk, then in principle the disk can adjust to an approximate steady
state with infall balanced by accretion. To fix ideas, we assume
that the disk beyond 1 AU is mostly heated by irradiation from the
central protostar of mass $M_{*}$, so that the temperature $T
\propto r^{-1/2}$. For a fully viscous disk, we adopt the usual
parametrization of the viscosity $\nu = \alpha c_s^2 /\Omega$, where
$c_s$ is the sound speed (for a molecular gas) and $\Omega$ is the
(roughly Keplerian) angular velocity.  Then \be t_{\nu} \sim {r^2
\Omega \over \alpha c_s^2} \sim 1.3 \times 10^3 \alpha_{-1}^{-1}\,
M_1\, T_{300}^{-1} \, R_{AU}\,  yr \,, \en where $\alpha_{-1} \equiv
\alpha/0.1$ is the viscosity parameter $M_1 \equiv M_{*}/M_\odot$,
$R_{AU} \equiv r/{\rm AU}$, and $T_{300}$ is the temperature at 1 AU
in units of $300$K. From this relation we see that a fully viscous
disk might be able to keep up with mass infall over typical
protostellar lifetimes of $\sim 10^5$ yr if the radius at which
matter is being added satisfies $R_{AU} \lesssim 10^2 \alpha_{-1}$.
In a layered disk picture, the viscosity may need to be modified
such that $\nu=\alpha c_{s} h$, where $h$ should be the thickness of
the active layer instead of the midplane scale height. However, this
difference is significant only if the temperatures differ at the
active layer and the midplane, which they do not unless there is
some midplane viscosity. In any case, Eq. (2) provides an upper
limit to $R_{AU}$. Since typical observational estimates of infall
radii are $\sim$10-100 AU \citep{1993ApJ...414..676K}, protostellar
infall to a constant $\alpha$ disk is likely to pile up unless
$\alpha$ is relatively large.

A more serious problem is that protostellar disks are unlikely to have
constant $\alpha$.  The best studied mechanism for angular momentum
transport in disks, turbulence driven by the MRI (e.g., Balbus \& Hawley
1998), requires a minimum ionization fraction to couple the magnetic
fields to the mostly neutral disk.  As substantial regions of
protostellar disks will generally be too cold for thermal (collisional)
ionization, ionization by nonthermal processes becomes important.  This
led Gammie (1996) to suggest a layered model in which non-thermally
ionized surface layers are magnetically coupled while the disk midplane
remain inert.  We modify Gammie's analysis by assuming that the heating
of the outer disk is not determined by local viscous dissipation but by
irradiation from the central protostar, as above.

The mass accretion rate in a layered disk is
\begin{equation}
{\mdot} = 6 \pi r^{1/2} {\partial \over \partial r} \left(
2 \Sigma_a \nu r^{1/2}\right) \,,
\end{equation}
where $\Sigma_a$ is the (one-sided) surface density of the active
layers.  Taking $\Sigma_a = $ constant, and assuming that the disk
temperature $T \propto r^{-1/2}$,
\begin{equation}
{\mdot} = 5 \times 10^{-7} \Sigma_{100} T_{300} \alpha_{-1} R_{AU}\,,
\label{eq:mdotlayer}
\end{equation}
where $\Sigma_{100} \equiv \Sigma_a/100 {\rm g\, cm^{-2}}$.

Our nominal value of $\alpha = 0.1$ may be reasonable for
well-ionized regions, but it may be an overestimate for the outer
regions of T Tauri disks (see \S 5.3).  Also, the fiducial value for
$\Sigma_a$ is based upon Gammie's (1996) assumption of cosmic ray
ionization, which may be an overestimate due to exclusion of cosmic
rays by scattering and advection in the magnetized protostellar
wind.  X-rays provide a higher ionization rate near the surface of
the disk but are attenuated more rapidly than cosmic rays (Glassgold
\& Igea 1999), yielding similar or smaller $\Sigma_a$.  Both
calculations assume that absorption of ions and electrons by grains
is unimportant, which is only true if small dust is highly depleted
in the active layer (e.g., Sano \etal 2000, also Ilgner \& Nelson
2006a,b,c).  In summary, it is likely that the estimate in equation
(\ref{eq:mdotlayer}) is an upper limit, and thus it appears unlikely
that the MRI can transport mass at $r \sim$ a few AU at protostellar
infall rates $2 \times 10^{-6} - 10^{-5} \msunyr$. MRI transport
resulting from non-thermal ionization might however move material
adequately in response to infall at $r \gtrsim 10 - 100$ AU.

On the other hand, if some nonmagnetic angular momentum transport
mechanism can get matter in to $r \lesssim 1$ AU, {\em thermal}
ionization can occur and activate the MRI.  A {\em minimum} disk
temperature is given by the effective temperature generated solely by
local energy dissipation
\begin{equation}
T > T_{eff} \sim 1600 ( M_1 \mdot_{-5} )^{1/4} \, (R / 0.2 AU)^{-3/4}\,
{\rm K}\,,
\end{equation}
where $\mdot_{-5} \equiv \dot{M}/10^{-5} \msunyr$.  Radiative trapping
in an optically thick disk will make internal temperatures even higher.
If $T \gtrsim 1400$ K most of the silicate particles will evaporate,
thus eliminating a major sink for current-carrying electrons.
Therefore, high accretion rates can potentially activate the MRI on
distance scales of order 1 AU or less.

If magnetic angular momentum transport is weak then mass will accumulate
in the disk until the disk becomes gravitationally unstable, at which
point gravitational torques can transfer mass inward.  GI alone may
result cause accretion outbursts (Vorobyov \& Basu 2006, 2008), although
the details of disk cooling are crucial in determining if such bursts
actually occur due to pure GI.  Moreover the GI may be unable to drive
accretion in the inner disk.  GI sets in when the Toomre parameter
\be
Q = {c_s \kappa \over \pi G \Sigma} \simeq {c_s \Omega \over \pi G
\Sigma}\sim 1 \,,
\label{eq:q}
\en
where we have set the epicyclic frequency $\kappa \simeq \Omega$,
appropriate for a near-Keplerian disk.  At small radius $\Omega$ and
$c_s$ will be large and therefore $\Sigma$ must also be large if we are
to have GI.  Since rapid accretion causes significant internal heating
(compared to heating by protostellar irradiation), large surface
densities imply significant radiative trapping, raising internal disk
temperatures above the effective temperature estimate above.  Thus, when
considering rapid mass transfer by GI, either in a quasi-steady state or
in bursts, it is necessary to consider thermal MRI activation in the
inner disk.

The above considerations suggest that the only way low-mass protostellar
disks can accrete steadily during infall is if a smooth transition can
be made from the GI operating on scales of $\sim 1-10$~AU to the
thermally-activated MRI at smaller radii.  To test this idea, we have
constructed a series of steady-state disk models with realistic
opacities.  We compute both MRI and GI steady models and then
investigate whether a smooth, steady, or quasi-steady transition is
likely.  Our results indicate that making the optimistic assumptions of
steady GI and MRI accretion results in a contradiction for infall rates
thought to be typical of low-mass protostars.

\section{Methods}

We compute steady disk models employing cylindrical coordinates ($r,z$),
treating radiative energy transport only in the vertical ($z$)
direction.  Energy conservation requires that
\begin{equation}
\sigma T_{eff}^4 = {3 G M_{*}\dot{M} \over 8\pi\sigma r^{3}}
\left(1-\left(\frac{r_{in}}{r}\right)^{1/2}\right)\,,
\label{eq:Fv}
\end{equation}
where $M_*$ is the central star's mass and we have assumed that the
disk is not
so massive as to make its rotation significantly non-Keplerian.
Balance between heating by dissipation of turbulence and radiative
cooling requires that
\begin{equation}
\frac{9}{4}\nu\rho_{z}\Omega^{2}=\frac{d}{d
z}\left(\frac{4\sigma}{3}\frac{d T^{4}}{d{\tau}}\right) \, ,
\label{eq:rt}
\end{equation}
where
\begin{equation}
\nu=\alpha c_s^2/\Omega \,
\label{eq:nu}
\end{equation}
and
\begin{equation}
d\tau = \rho \kappa dz \, ,
\label{eq:tau}
\end{equation}
and $\kappa$ is the Rosseland mean opacity.
We have updated the fitting formulae provided by Bell \& Lin (1994) for
the Rosseland mean opacity to include more
recent molecular opacities and an improved treatment of the
pressure-dependence of dust sublimation (Zhu \etal 2007). The new
fit and a comparison with the Bell \& Lin (1994) opacity treatment is given in
the Appendix.

Convection has not been included in our treatment.
\cite{linpapa1980} show that for a power law opacity
($\kappa=\kappa_{0}T^{\beta}$), convection will occur when
$\beta\gtrsim 1$.  Our opacity calculations show that $\beta\gtrsim
1$ only occurs for $T\gtrsim 2000 K$. As our steady-state analysis
depends upon disk properties for T $\lesssim 1400$~K, the neglect of
convection will not affect our results (see also Cassen 1993).

We ignore irradiation of the disk by the central star, as we are
assuming high accretion rates and a low central protostellar luminosity.
The diffusion approximation (equation \ref{eq:rt}) is adequate since the
disk is optically thick at the high mass accretion rates
($\dot{M}>10^{-7}M_{\odot}/yr$) we are interested in.

We also require hydrostatic equilibrium perpendicular to the disk plane,
\begin{equation}
\frac{dP_{z}}{dz}=\frac{GM_{*}\rho z}{r^{3}}  \, ,
\label{eq:hydro}
\end{equation}
and use the ideal gas equation of state
\begin{equation}
P=\frac{k}{\mu}\rho T \,. \label{eq:EOS}
\label{eq:eos}
\end{equation}
Given a viscosity prescription, equations (\ref{eq:q}) - (\ref{eq:eos})
can be solved iteratively for the vertical structure of the disk at each
radius, resulting in self-consistent values of the surface density
$\Sigma$, and the temperature at the disk midplane $T_c$.

In detail, we use a shooting method based on a Runge-Kutta integrator
rather than a relaxation method (e.g., D'Alessio \etal 1998) to solve
the two-point boundary value problem.  Given $\alpha$ and $\dot{M}$ at
$r$, we fix $z = z_i$ and set $T = T_{eff}$ and $\tau = 2/3$ (this is
adequate in the absence of significant protostellar irradiation), then
integrate toward the midplane.  We stop when the total radiative flux $=
\sigma T_{eff}^4$ at $z = z_f$.  In general $z_f \ne 0$; we alter the
initial conditions and iterate until $z_f = 0$.

For an MRI active disk we fix $\alpha = \alpha_M$, assuming the disk
is active through the entire column.
We then check to see if thermal ionization is sufficient
or if the surface density is low
enough that non-thermal ionization is plausible.  The exact
temperatures above which MRI activity can be sustained are somewhat
uncertain; here we assume the transition occurs for a central temperature of 1400~K,
when the dust grains that can absorb ions and electrons and thus
inactivate the MRI (e.g., Sano \etal 2000) are evaporated.  We set
$\alpha_M = 10^{-2}$ to $10^{-1}$ to span a reasonable range given
current estimates (see \S 5.3).

For simplicity we neglect the possible presence of an actively accreting,
non-thermally-ionized layer.  This omission will not affect our results at
high accretion rates, for which the layered contribution is unimportant
(equation \ref{eq:mdotlayer}); our approximation then breaks down for
$\mdot \le 10^{-6} \msunyr$ for large values of $\Sigma_a$ and $\alpha_M$.

For the steady GI disk models $\alpha$ is not fixed.  Instead we start
with a large value of $\alpha = \alpha_Q$ and then vary $\alpha_Q$ until
$Q = 2$.  The adoption of the local treatment of GI energy dissipation
requires some comment.  Since gravity is a long-range force a local
viscous description is not generally applicable (Balbus \& Papaloizou
1999). However, as Gammie (2001) and Gammie \& Johnson (2003) argue, a
local treatment is adequate if $\lambda_c \equiv 2 c_s^2/(G \Sigma) =
2\pi H Q \lesssim r$; here $\lambda_c$ is the characteristic wavelength
of the GI.  More broadly, our main result involves order-of-magnitude
arguments; that is, as long as inner disks must be quite massive to
sustain GI transport, and as long as there is {\em some} local
dissipation of energy as this transport and accretion occurs, steady
accretion will not occur for a significant range of infall rates. To
change our conclusions dramatically, one would need to show that the GI
causes rapid accretion through the inner disk without substantial local
heating.  We return to this issue in \S 5.1.

\section{Results}

Figures 1a-d show steady disk results for a central star mass of $1
\msun$ and accretion rates of $10^{-4}$, $10^{-5}$, $10^{-6}$, and
$10^{-7} \msunyr$.  Proceeding counterclockwise from upper left, the
panels show the central disk temperature, $\alpha_Q$, the one-sided
surface density $\Sigma \equiv \int_0^\infty dz \rho$, and the viscous
timescale $r^2/\nu$ as a function of radius. The solid curves show
results for pure-GI models, while the dotted and dashed curves show
results for $\alpha_M = 0.1$ and $0.01$, respectively.

The upper left panels show that the central temperatures rise more
dramatically toward small radius in the GI models than in the MRI
models.  The GI models have higher temperatures because their higher
surface densities lead to stronger radiative trapping.  The GI solutions
in these high-temperature regimes are unrealistic because they assume
the MRI is absent, when it seems likely the MRI will in fact be active.
These high temperature states do, however, suggest the possibility of
thermal instability in the inner disk at high accretion rates,
especially as the solutions near $\sim 3000$~K represent unstable
equilibria (e.g., Bell \& Lin 1994; \S 5).  We consider the MRI models
to be inconsistent at $T \lesssim 1400$ K (collisional ionization would
be absent) and when $\Sigma > \Sigma_a < 100 {\rm g\, cm^{-2}}$.

Can a smooth or steady transition between MRI and GI transport
occur? The transition region would be the ``plateau'' in the
temperature structure which occurs near $T \sim 1400$~K (see Figure
1).  This plateau is a consequence of the thermostatic effects of
dust opacity, which vanishes rapidly at slightly higher
temperatures.  A small increase in temperature past this critical
temperature causes a large decrease in the disk opacity and thus the
optical depth; this in turn reduces the radiative trapping and
decreases the central temperature. Thus disk models tend to hover
around the dust destruction temperature over roughly an order of
magnitude in radius, with the plateau occurring farther out in the
disk for larger accretion rates.  Since the plateau is connected
with the evaporation of dust, it corresponds to a region where we
might expect MRI activity.\footnote{There will be hysteresis because
the dust size spectrum in a parcel of gas will depend on the
parcel's thermal history.  Heating the parcel destroys the dust and
the accumulated effects of grain growth.  Cooling it again would
presumably condense dust with small mean size (and therefore a
strong damping effects on MHD turbulence).  The opacity would then
vary strongly with time as the grains grow again.  These effects are
not considered here.}

First consider the case $\mdot = 10^{-4} \msunyr$ (upper left corner of
Figure 1).  The plateau region is very similar in extent for all models.
More importantly, $\alpha_Q \sim 10^{-2}$ in this region, and so the
surface densities of the GI and $\alpha_M = 10^{-2}$ models are nearly
the same.  This suggests that a steady disk solution is plausible with a
transition from GI to MRI at a few AU for these parameters.  Depending
on the precise thermal activation temperature for the MRI, a smooth
transition at around 10 AU might also occur for $\alpha_M = 0.1$.

Next consider the case $\mdot = 10^{-5} \msunyr$ (upper right corner of
Figure 1).  Here $\alpha_Q \sim 10^{-3}$ in the plateau region, with
resulting surface densities much higher than for either of the MRI
cases.  This discrepancy in $\alpha$ and $\Sigma$ between the two
solutions makes a steady disk unlikely.  A small increase in surface
density in a GI model near the transition region, resulting in increased
heating and thus thermal activation of the MRI, would suddenly raise the
effective transport rates by one or two orders of magnitude, depending
upon $\alpha_M$.  The result would be an accretion outburst.  This is
qualitatively the same situation as proposed for outbursts in dwarf
novae, where thermal instability is coupled to an increase in $\alpha$
from the initial low state to the high state (similar to what Bell \&
Lin 1994 adopted to obtain FU Ori outbursts).  Our inference of
non-steady accretion also agrees with the time-dependent one-dimensional
models of Armitage \etal (2001) and of Gammie (1999) and Book \&
Hartmann (2005), as discussed further in in \S 5.

A similar situation holds at $10^{-6} \msunyr$, although the
evolutionary (viscous) timescales of the GI model are of order
$10^5$~yr, comparable to protostellar infall timescales.  At this infall
rate, the disk would only amass $\sim 0.1 \msun = 0.1 M_*$, and so the
disk might not need to transfer this mass into the star to avoid GI.  At
$10^{-7} \msunyr$, evolutionary timescales become much longer than
protostellar lifetimes, and become comparable to T Tauri lifetimes; disk
material can pile up without generating GI transport and consequent
thermal activation of the MRI.  In addition, an $\alpha_M = 0.1$ value
could result in a steady disk with surface densities low enough to be
activated entirely by cosmic ray or X-ray ionization.  This does not
mean, however, that T Tauri disks do not have layered accretion, as the
surface density distribution depends upon the history of mass transport.

The results of our calculations are summarized in the $\mdot - r$
plane in Figure 2.  The solid curves farthest to the lower right,
labeled $R_Q$, are the radii at which the pure GI-driven disk would
have a central temperature of 1400 K (at which temperature the dust
starts to sublimate), and thus activate the MRI.  Moving up and
left, the solid curve labeled $R_{M}$ denotes the radii at which a
pure MRI disk of the given $\alpha_M$ would have a central
temperature of 1400 K.  When these two curves are close together, or
cross, $\alpha_Q$ and $\alpha_M$ are similar, making possible a
smooth transition between GI and MRI and thus steady accretion.  In
the (shaded) regions between these two curves the viscosity
parameters diverge, making non-steady accretion likely.

The radial regions at which we predict material will pile up, trigger
the MRI, and result in rapid accretion lie in the shaded regions. The
dotted curve shows $R_{Q}$ and $R_{M}$ where the disk has a central
temperature of 1800 K (at which temperature all dust has sublimated).
$R_{Q}$ and $R_{M}$ at 1800 K are smaller than they are at 1400 K
because of the plateau region discussed above.  Thus if the MRI trigger
temperature is higher the outbursts are expected to be shorter because
the outburst drains the smaller inner disk ($r < R_{Q}$) on the viscous
timescale.

Figure 2 indicates that non-steady accretion, with potential outbursts,
is predicted to occur for infall rates $\lesssim 10^{-5} \msunyr$ for
$\alpha_M = 0.01$ and $\lesssim 10^{-4} \msunyr$ for $\alpha_M = 0.1$.
As described above, for $\mdot < 10^{-6} \msunyr$ outbursts are
unlikely, simply because the transport timescales are too long.
Outbursts are expected to be triggered at $r \sim 1-10$~AU for
protostellar infall rates $\sim 10^{-5} - 10^{-6} \msunyr$.  These
predictions are relatively insensitive to the precise temperature of
MRI activation; the dotted curves in Figure 2 show the results for a
critical MRI temperature of 1400 K, which simply shift the regions of
instability to slightly smaller radii without changing the qualitative
results.

The other shaded band in Figure 2 denotes the region where thermal
instability might occur.  The two limits correspond to the two limiting
values of the ``S curve'' (e.g., Faulkner, Lin, \& Papaloizou 1983) at
which transitions up to the high (rapid accretion) state and the low
(slow accretion) state occur.

Figures 3-6 show results for central star masses of $0.3$ and $0.05
\msun$, respectively.  The predictions are qualitatively similar to the
case of the $1 \msun$ protostar, with the exception that thermal
instability is less likely for the brown dwarf.  This also implies
generally unstable protostellar accretion for more massive protostars
during the time that they are increasing substantially in mass.

Much of the overall behavior of our results derive from the general
property that disk temperatures rise strongly toward smaller radii.  For
optically-thick viscous disks, the central temperatures are proportional
to
\be
T_c \sim T_{eff} \tau^{1/4} \propto \mdot^{1/4} r^{-3/4} (\kappa_R
\Sigma)^{1/4}\,,
\en
where $\tau$ is the vertical optical depth.  Thus, even changes in
surface density for differing values of $\alpha$ result in modest
changes in radii where a specific temperature is achieved.  Changing the
mass accretion rate has a bigger effect, because $\Sigma \propto \mdot$.

\section{Discussion}

Our prediction of unsteady accretion during protostellar disk
evolution is the result of the inefficiency of angular momentum
transport of the two mechanisms considered here: the MRI, because of
low ionization in the disk; and the GI, because it tends to be
inefficient at small radii, where $\Omega$ and $c_s$ will be large,
forcing $\Sigma$ to be large. To provide a feeling of just how large
the surface density must be for $Q=2$ in the inner disk, at
accretion rates of $10^{-4}$ and $10^{-5} \msunyr$ for the $1 \msun$
star the disk mass interior to 1 AU would have to be $\sim 0.6
\msun$ and $\sim 0.5 \msun$, respectively (Fig. 7), which are
implausible large. At some point the disk must accrete most of its
mass into the star, forcing the inner disk temperatures to be very
large and thermally activating the MRI, resulting in outburst of
accretion. Here we consider whether the assumptions leading to this
picture are reasonable, then discuss applications to outbursting
systems.

\subsection{Outbursts?}

Our inference of cycles of outbursts of accretion - piling up of mass by
GI transport, followed by thermal triggering of the MRI - was found in
the models of Armitage \etal (2001), as well as in the calculations of
Gammie (1999) and Book \& Hartmann (2005).  We have also found
outbursting behavior in time-dependent two-dimensional disk models, to
be reported in a subsequent paper (Zhu, Hartmann, \& Gammie 2009).  Here
we compare our results with those of Armitage \etal.

Figure 8 shows the results of our stability calculations for
parameters and opacities adopted by Armitage \etal: a central star
mass of $1 \msun$, $\alpha_M = 0.01$, and an assumed triggering
temperature for the MRI of 800 K. Armitage \etal found steady
accretion at an infall rate of $\mdot = 3 \times 10^{-6} \msunyr$
but outbursting behavior at $1.5 \times 10^{-6} \msunyr$.  This is
reasonably consistent with our calculations; $R_M$ and $R_Q$ are
close together at $\mdot = 3 \times 10^{-6} \msunyr$ and cross near
$\mdot = 1 \times 10^{-5} \msunyr$, suggesting stable accretion
somewhere in this range. Armitage \etal find that the MRI is
triggered at about 2 AU, whereas our analysis (for $\mdot \sim
10^{-6} \msunyr$) would suggest a triggering radius of about 3 AU.
Our ability to reproduce the results of Armitage et al. is adequate,
considering that steady models do not precisely reproduce the
behavior of time-dependent models, and that the form of $\alpha_Q$
used by Armitage \etal is somewhat different from ours, though it
still retains the feature of non-negligible GI only for small $Q$.

Our finding of non-steady accretion is the result of assuming no other
significant level of angular momentum transport that is not due to GI or
thermal MRI.  \cite{terquem2008} has shown that steady accretion is possible
for a layered disk accreting at $\dot{M}=10^{-8}\msun/yr$
if there is a non-zero (non-gravitational) viscosity in disk regions
below the surface active layers.  Simulations have indicated that
active layers can have an effect on non-magnetically active regions
below, producing a Reynolds stress promoting accretion
in the lower regions \citep{fleming2003,turner2008,2008A&A...483..815I}.
We argue that this effect is unlikely to be important
for the much higher accretion rates considered here, simply because the
amount of mass transfer that needs to occur is much higher than what is
sustainable by a non-thermally ionized surface layer.  It seems implausible
that a small amount of surface energy and turbulence generation can activate
a very large amount of turbulence and energy dissipation in a much more massive
region.

\subsection{Local vs. non-local GI transport}

We have adopted a local formalism for GI whereas it has non-local
properties.  Furthermore, we have adopted azimuthal symmetry in
calculating the dissipation of energy whereas energy will be deposited
in nonaxisymmetric spiral shocks.  Neither of these assumptions is
strictly correct.

Boley \etal (2006) performed a careful analysis of the torques in a
three-dimensional model of a self-gravitating disk, including radiative
transfer.  They found that the mass transfer was dominated by global
modes, but could be consistent with a locally-defined $\alpha(r,t)$.
This result did not hold near the inner and outer edge of their disk,
although this is not surprising as these regions were characterized by
$Q > 2$ and thus one would not expect the GI to be operating.  Boley
\etal were unable to address whether energy dissipation was localized.
Nevertheless it is difficult to imagine that gravitational instability
could avoid some heating in regions with $Q \sim 1$, and only relatively
small amounts of heating are required to activate the MRI at small
radii.

The details of the disk temperature structure near 1 AU must be found by
three dimensional simulations of the GI with realistic cooling.  The
analysis presented here suggests that pure GI in the absence of MRI
tends to lead to very long transport times in the inner disk, as
required by our low values of $\alpha_Q$.  This presents two potential
technical problems for a numerical investigation: first, numerical
viscosity must be smaller than $\alpha_Q$ to follow the evolution; and
second, the disk must be followed over long, evolutionary timescales.
It will be challenging to follow the GI near 1 AU numerically.

\subsection{What is $\alpha_M$?}

The magnetic transport rate $\alpha_M$ is constrained by both
observations and theory.  A recent review of the observational evidence
by King, Pringle, \& Livio (2007) argues that $\alpha_M$ must be large,
of the order 0.1-0.4, based in part on observations of dwarf novae and
X-ray binaries where there is no question of gravitational instability.
Our own analysis required $\alpha \sim 0.1$ in FU Ori (Zhu \etal 2007).

On the theoretical side the situation is murky.  Early calculations
\citep{hgb96} suggested that for ``shearing box'' models with zero mean
azimuthal and vertical field $\alpha_M \simeq 0.01$.  Recent work
\citep{fp07a}, however, shows that $\alpha_M$ does not converge in the
sense that $\alpha_M \rightarrow 0$ as the numerical resolution
increases.

But are the zero mean field models relevant to astrophysical disks?
Global disk simulations \citep{hkvh05, mn07, bhk08}, local disk
simulations in which the mean field is allowed to evolve because of
the boundary conditions \citep{bran95}, and observations of the
galactic disk \citep{va04} all exhibit a ``mean'' azimuthal field
when an average is taken over areas of $\gtrsim H^2$ in the plane of
the disk.  This suggests that the zero mean field local models are a
singular case, and that mean azimuthal field models are most
relevant to real disks (strong vertical fields would appear to be
easily removed from disks according to the plausible
phenomenological argument originally advanced by \cite{vanball}).

So what do numerical simulations tell us about disks with mean
azimuthal field?  Recent work shows that in this case the outcome
depends on the magnetic Prandtl number $Pr_M \equiv \nu/\eta$
\citep{fp07b, ll07} ($\nu \equiv$ viscosity and $\eta \equiv$
resistivity) and that $\alpha_M$ is a monotonically increasing
function of $Pr_M$.  This intriguing result, and the fact that YSO
disks have $Pr_M \ll 1$ throughout (although more dimensionless
parameters are required to characterize YSO disks, where the Hall
effect and ambipolar diffusion can also be important), might suggest
that $\alpha_M$ should be small.  But the numerical evidence also
shows that $\alpha_M$ depends on $\nu$ in the sense that the
dependence on $Pr_M$ weakens as $\nu$ decreases.  In sum, the
outcome is not known as $\nu$ drops toward astrophysically plausible
values.  Mean azimuthal field models with effective $Pr_M \sim 1$
\cite{ggsj} are also not fully converged; they show that $\alpha_M$
{\it increases}, albeit slightly, as the resolution is increased.
For a mean field with plasma $\beta = 400$ \cite{ggsj} find
$\alpha_M = 0.03$ at their highest resolution. In disks with an
initial strong azimuthal magnetic field in equipartition with
thermal pressure,
 \cite{2008A&A...490..501J} find $\alpha=0.1$ resulting from a combination of the
Parker instability and an MRI-driven dynamo.

Very small $\alpha_M$ would pose a problem for T Tauri accretion.  In
the layered disk model, Gammie estimated the accretion rate to be
\begin{equation}
\mdot~ \sim~ 2 \times 10^{-8} \left ( {\alpha_M \over 0.01} \right )^2
\,
\left ( {\Sigma_a \over 100 {\rm g \, cm^{-2}} } \right )^3
\msunyr\,,
\end{equation}
where $\Sigma_a$ is the surface density of the layer which is
non-thermally ionized.  Thus with $\alpha_{M} \lesssim 10^{-3}$ it
would be difficult to explain typical T Tauri accretion rates.

On the other hand $\alpha_M \sim 0.1$ could cause the outer disks of T
Tauri stars to expand to radii of 1000 AU or more in 1 Myr (Hartmann
\etal 1998).  There is no particular reason why the $\alpha_M \sim 0.1$
that we estimated for the thermally-ionized inner disk region in FU Ori
should be the same as the effective $\alpha$ in the outer disks of T
Tauri stars, which cannot be thermally ionized.

\subsection{Protostellar accretion}

Our models predict that most low-mass protostars will be accreting more
slowly than matter is falling onto their disks.  This is consistent
with observational results, as outlined in the Introduction.  The
results of Armitage \etal (2001) suggested that steady accretion might
be possible at $\sim 3 \times 10^{-6} \msunyr$ and above (for $1
\msun$).  We find a different result because we adopt a significantly
higher temperature for thermal MRI activation, closer to that required
for dust evaporation.  This means that our MRI triggering occurs at
smaller radii, where the GI is less effective.  It does seem likely that
higher activation temperatures than the 800 K adopted by Armitage \etal
are more plausible.  Even if thermal ionization in the absence of dust
is sufficient at around 1000 K in statistical equilibrium, ionization
rates are so low that equilibrium is unlikely (e.g.  Desch 1998).  We
also note that Armitage \etal were unable to obtain the high accretion
rates and short outburst durations characteristic of FU Ori objects, but
Book \& Hartmann (2005) were able to reproduce the FU Ori
characteristics better with a higher MRI activation temperature.

At infall rates $\gtrsim 10^{-4} \msunyr$, our models predict (quasi-)
steady accretion (also Armitage \etal 2001); but such high rates are not
expected to last long, perhaps only during an initial rapid phase of
infall (Foster \& Chevalier 1993; Hartmann \etal 1994; Henriksen, Andre,
\& Bontemps 1997).  Testing this prediction may be difficult as
relatively few objects will be caught in this phase and they will likely
be heavily embedded.

At lower infall rates, GI-driven accretion timescales are longer than
evolutionary times and/or layered MRI turbulence may produce sufficient
mass transport.  Thus, we would not expect outbursts for Class II (T
Tauri) stars.

\subsection{FU Ori outbursts}

In our radiative transfer modeling of the outbursting disk system FU
Ori (Zhu \etal 2007), we found that to fit the {\em Spitzer Space
Telescope} IRS spectrum the rapidly-accreting, hot inner disk must
extend out to $\sim 1$~AU, inconsistent with a pure thermal instability
model.  In contrast, the results of this paper suggest thermal MRI
triggering can occur at a few AU, in much better agreement with
observation.

Our recent analysis of the silicate emission features of FU Ori (Zhu
\etal 2008) also suggests that the disk becomes dominated by irradiation
rather than internal heating  at distances of $\gtrsim 1$~AU, but this
is consistent with the results of this paper, as irradiation from the
central disk can dominate local viscous dissipation if the disk is
sufficiently flared.

We also found that the decay timescales of FU Ori suggest $\alpha_M \sim
10^{-1}$; large values of $\alpha_M$ are more likely to lead to
outbursting behavior.   High inner disk accretion rates also make
thermal instability more likely very close to the central star; the
presence or absence of this instability may account for the difference
in rise times seen in some FU Ori objects (Hartmann \& Kenyon 1996).

\section{Conclusions}

Our study predicts that the disk accretion of low-mass protostars
will generally be unsteady for typical infall rates.  During the
protostellar phase, GI is likely to dominate at radii beyond 1 AU
but not at smaller radii; in contrast, rapid accretion should drive
thermal activation of the MRI in the inner disk.  Because of the
differing transport rates comparable to typical infall values
results in high inner disk temperatures sufficient to trigger the
MRI.  This is a general conclusion, though if the external disk
accretion is driven by GI, the radius at which the MRI can be
triggered thermally is much larger, because of the high surface
density needed to produce a low value of $Q$.  Furthermore,
GI-driving in the inner disk results in a low value of $\alpha_Q$,
much lower than the expected $\alpha_M$, for a wide range of
$\mdot$.  The feature of mass accumulation at low external $\alpha$
followed by a change to a high inner viscosity is similar to thermal
instability models (and also Armitage \etal 2001).  Thermal
instabilities may also occur in the inner disk at very high
accretion rates, enhancing the potential for non-steady protostellar
accretion.

\appendix

\section{Appendix: Rosseland mean opacity}

The Bell \& Lin (1994) Rosseland mean opacity fit has been widely used
to study high temperature accretion disks (CV objects, FU Ori, \etal)
for more than a decade, with opacities generated almost two decades ago.
Our understanding of opacity sources (especially dust and molecular line
spectra) has improved both observationally and theoretically since
then \citep{alexander94,ferguson05,dalessio1998,dalessio2001,zhu2007}.

We have generated Rosseland mean opacity assuming LTE for a wide range
of temperature and pressure during our study of FU Orionis objects
\citep{zhu2007,2008arXiv0806.3715Z}. The molecular, atomic, and ionized
gas opacities have been calculated using the Opacity Distribution
Function (ODF) method \citep{kurucz04, kurucz042,castelli05,zhu2007}
which is a statistical approach to handling line blanketing when
millions of lines are present in a small wavelength range
\citep{kurucz74}. The dust opacity was derived by the prescription in
\cite{dalessio2001} \citep{2008arXiv0806.3715Z}. Our opacity has been
used not only to study FU Orionis objects but also to fit the gas
opacity for Herbig Ae star disks constrained by interferometric
observations \citep{2008arXiv0808.1728T}. The opacities are shown in
Figure \ref{fig:opa}). Compared with \cite{alexander94} or Zhu \etal
(2007,2008), the Bell \& Lin opacity lacks water vapor and TiO opacity
around 2000 K and has a lower dust sublimation temperature.

We have made a piecewise power-law fit to the Zhu \etal (2007, 2008)
opacity (analogous to the Bell \& Lin fit) to enhance computational
efficiency (Table \ref{fitopa}; see also Figure \ref{fig:opa}).  This
speedup has been useful in performing the calculations of this paper,
and is essential for our forthcoming two-dimensional hydrodynamic
simulations of FU Ori outbursts (Zhu, Hartmann, \& Gammie 2009).

\acknowledgments

We acknowledge useful conversations with Ken Rice and Dick Durisen.
This work was supported in part by NASA grant NNX08A139G, by the
University of Michigan, and by a Sony Faculty Fellowship, a Richard and
Margaret Romano Professorial Scholarship, and a University Scholar
appointment to CG.

\begin{figure}
\includegraphics[width=0.42\textwidth]{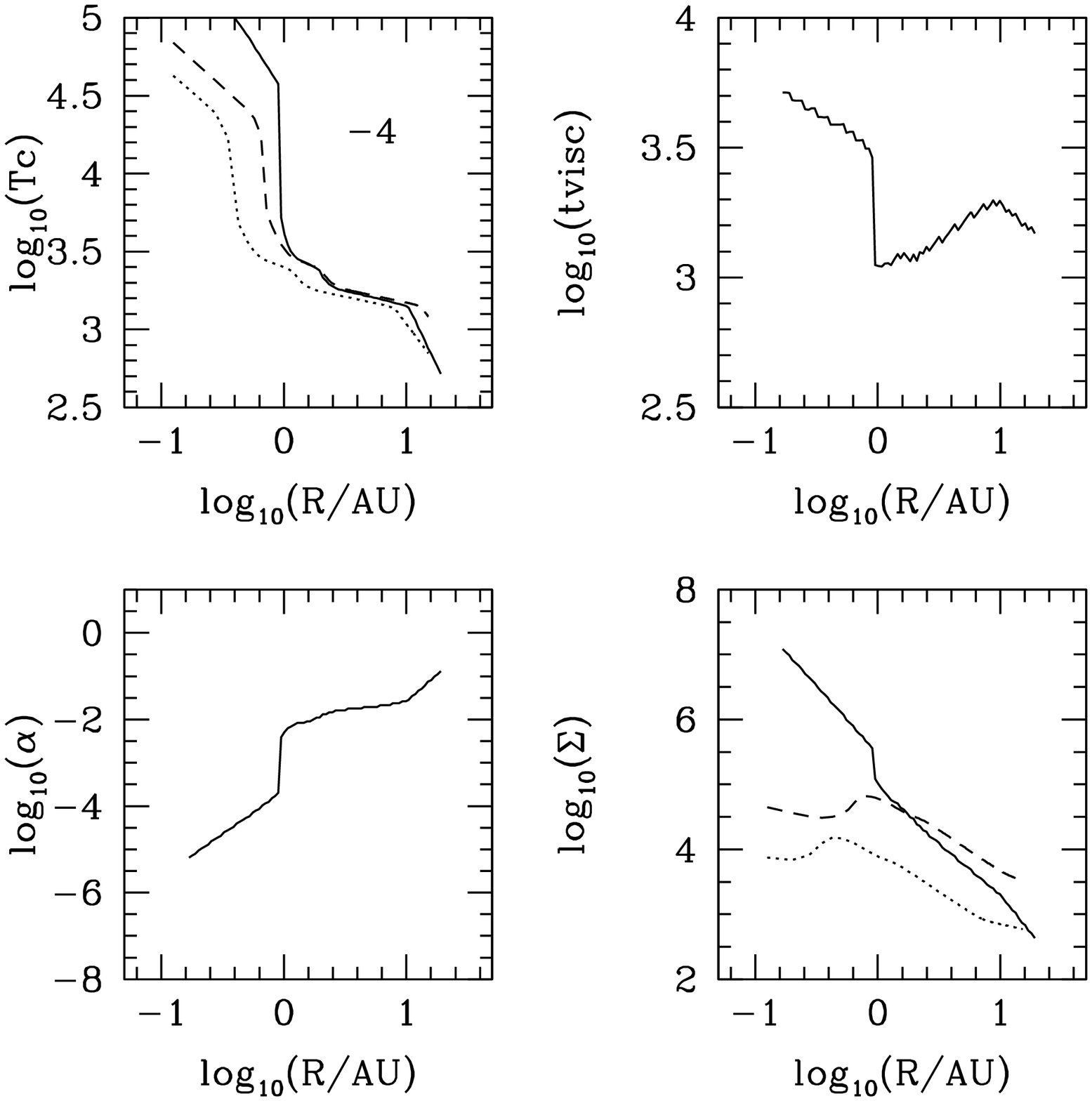} \hfil
\includegraphics[width=0.42\textwidth]{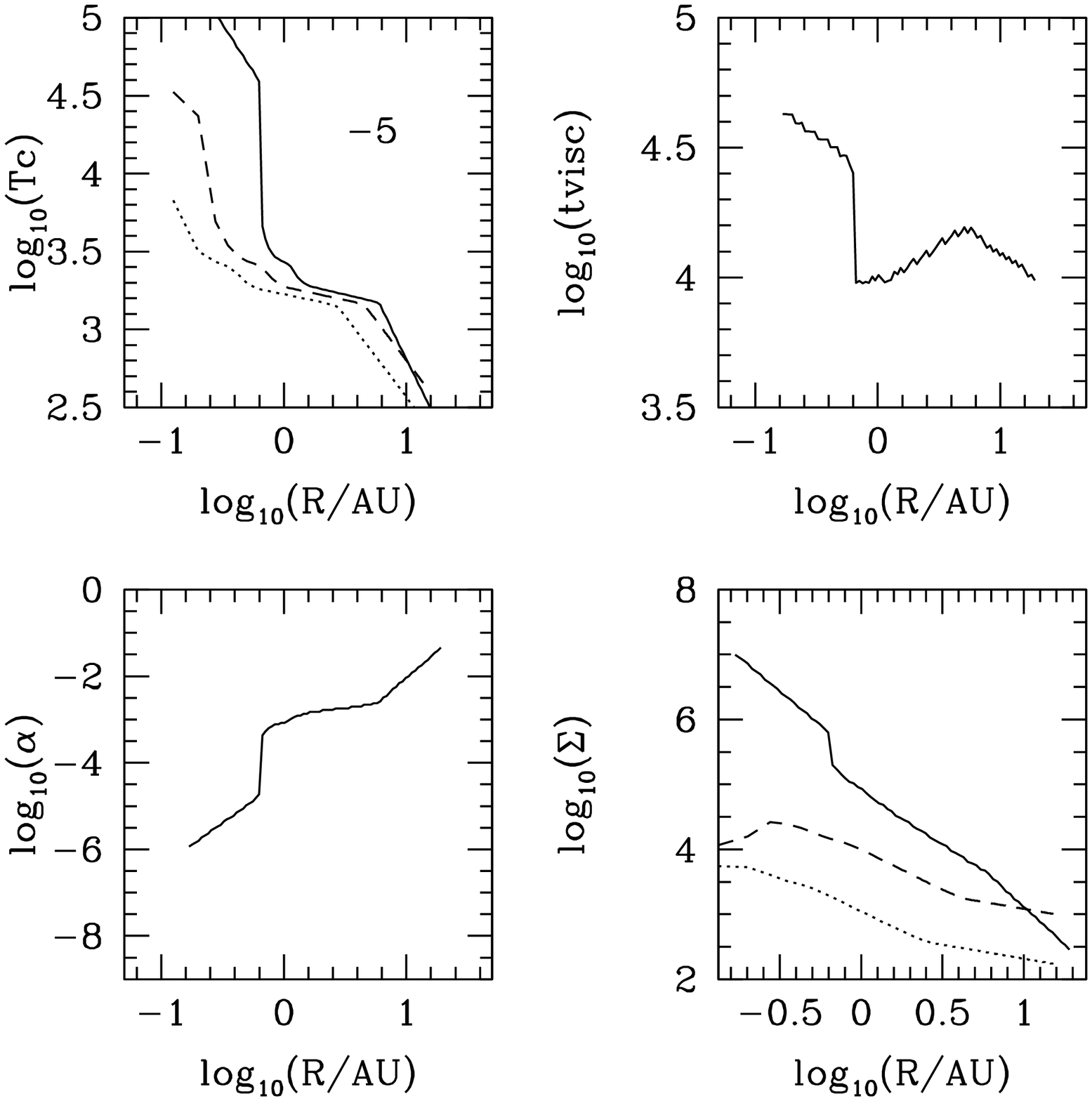} \\
\vskip 0.5cm
\includegraphics[width=0.42\textwidth]{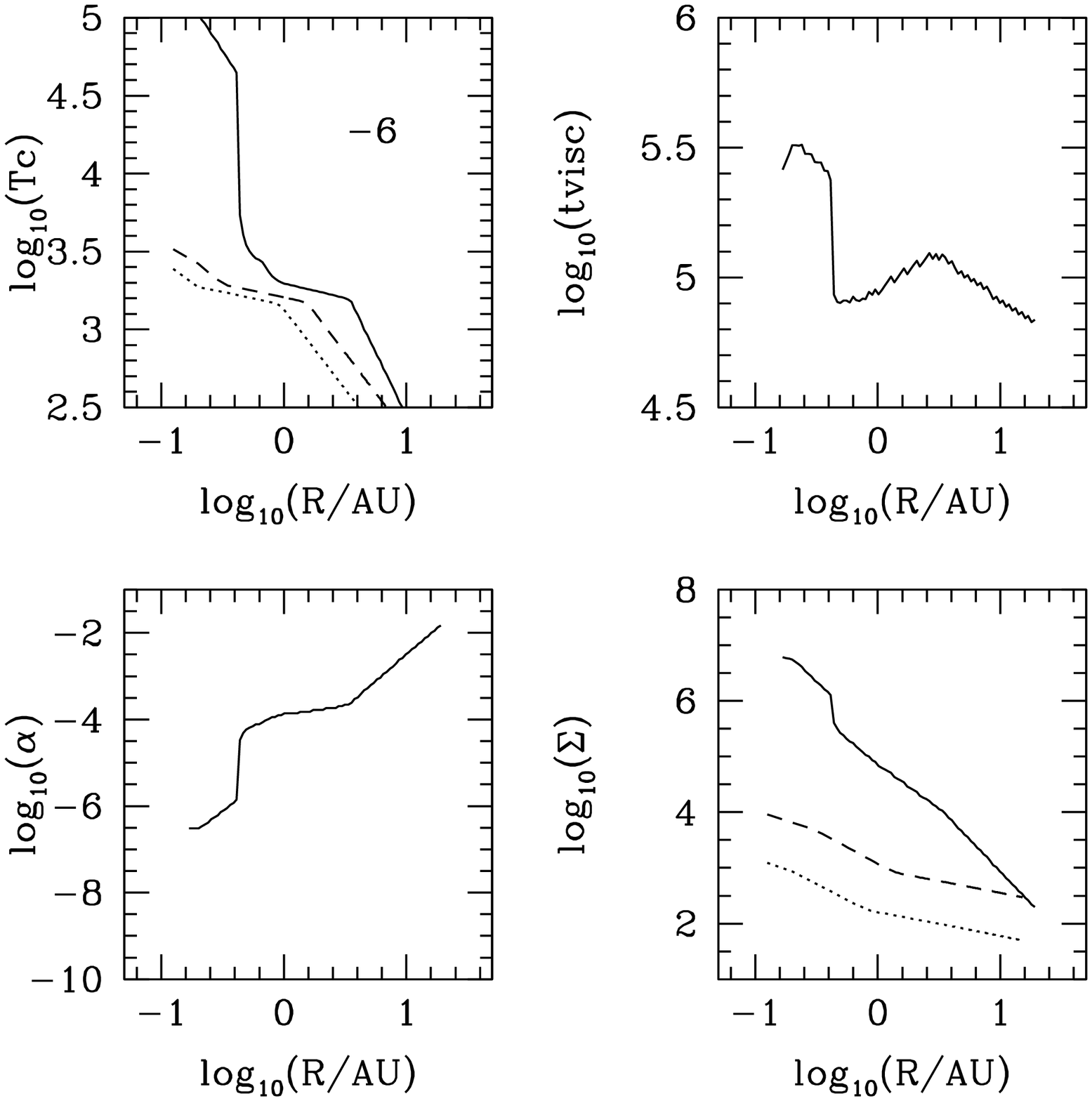} \hfil
\includegraphics[width=0.42\textwidth]{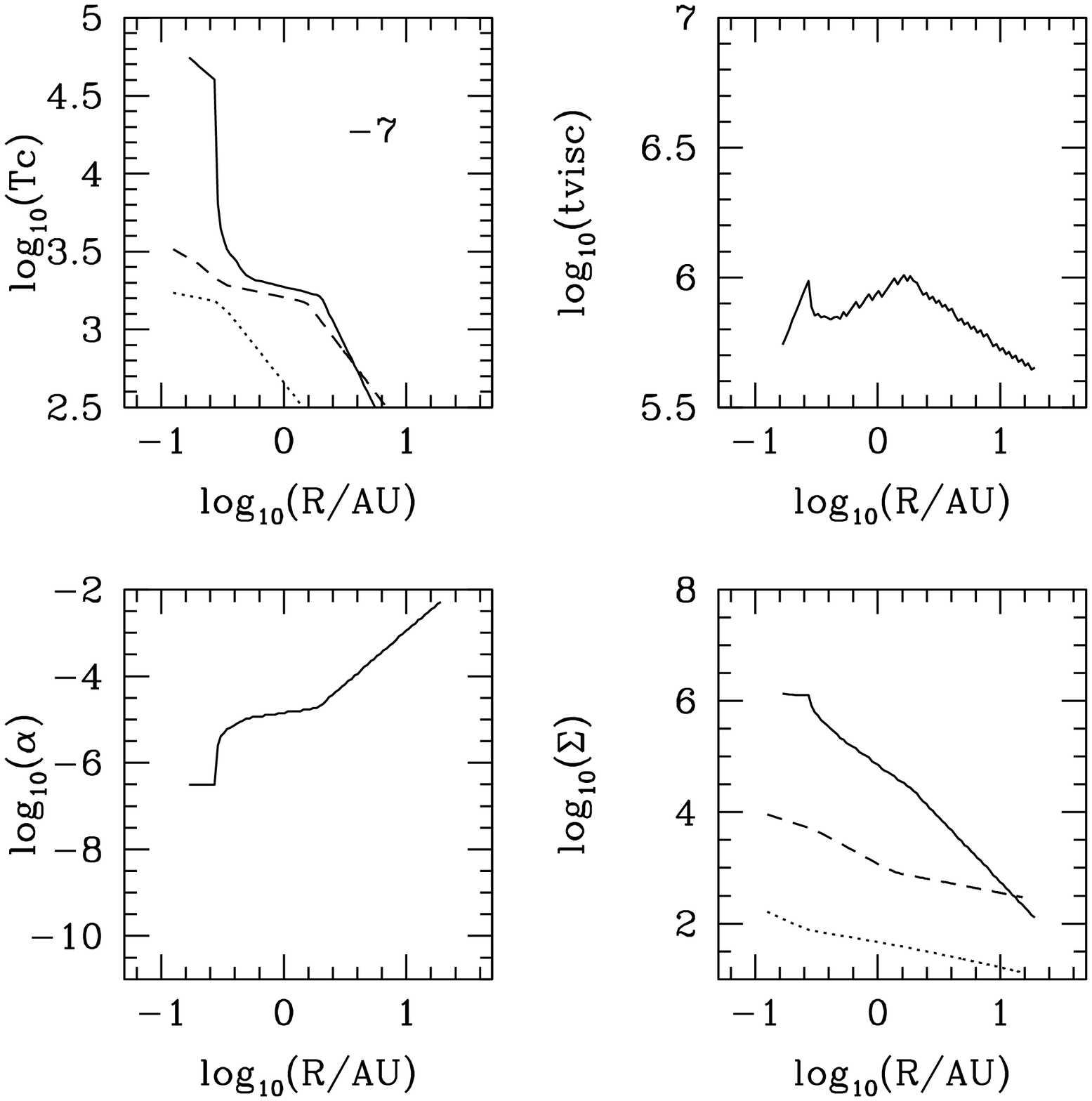} \\
\caption{Steady-state disk calculations for four accretion rates -
$10^{-4}$, $10^{-5}$, $10^{-6}$, and $10^{-7} \msunyr$, assuming a
central star of mass $1 \msun$.  The solid curves show solutions for
GI-driven accretion, as described in the text.  The dashed and dotted
curves yield results for steady disk models with a constant $\alpha
= 10^{-2}$ and $10^{-1}$, respectively (see text)}
\label{fig:f1}
\end{figure}

\begin{figure}
\includegraphics[width=0.42\textwidth]{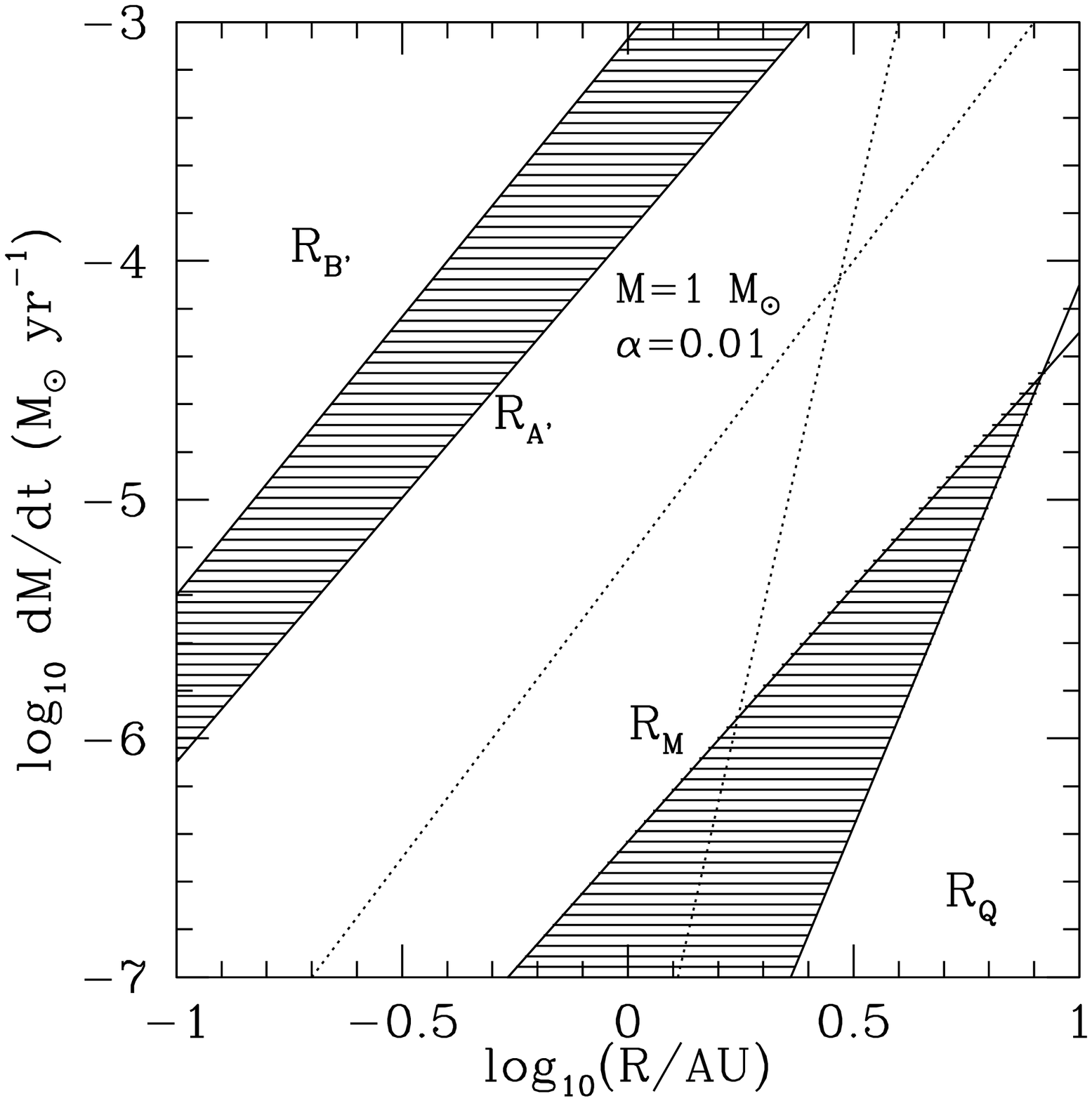} \hfil
\includegraphics[width=0.42\textwidth]{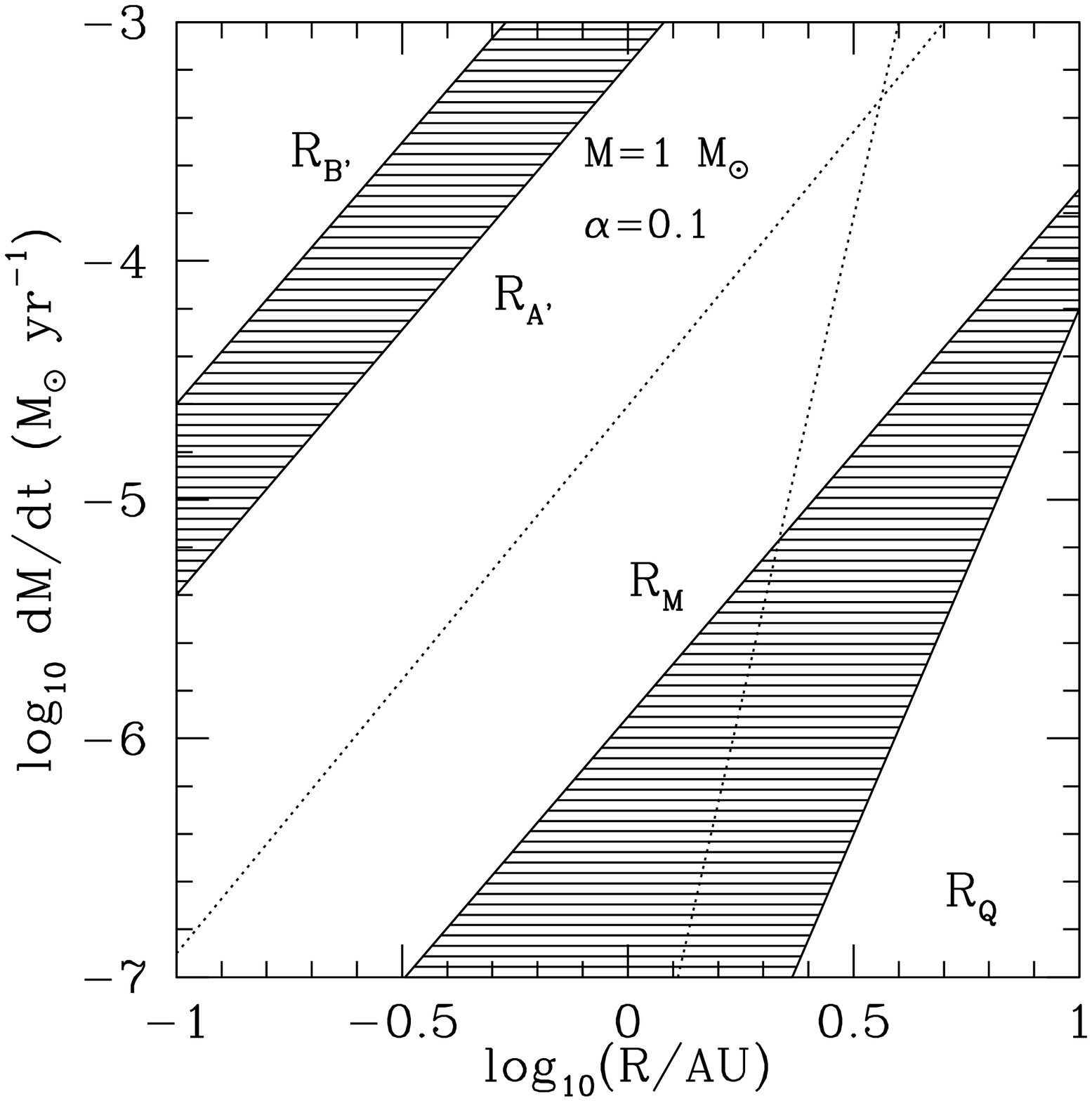} \\
\caption{
Unstable regions in the $r-\dot{M}$ plane for a $1 \msun$ central star.
The shaded region in the lower right shows where the central temperature
of steady GI models exceeds an assumed MRI trigger temperature of $1400$
K.  The dotted curves show $R_M$ and $R_Q$ (the boundaries of the shaded
region; see text for definition) for an MRI trigger temperature of
$1800$ K.  The shaded region in the upper left shows the region subject
to classical thermal instability.}
\label{fig:f2}
\end{figure}

\begin{figure}
\includegraphics[width=0.42\textwidth]{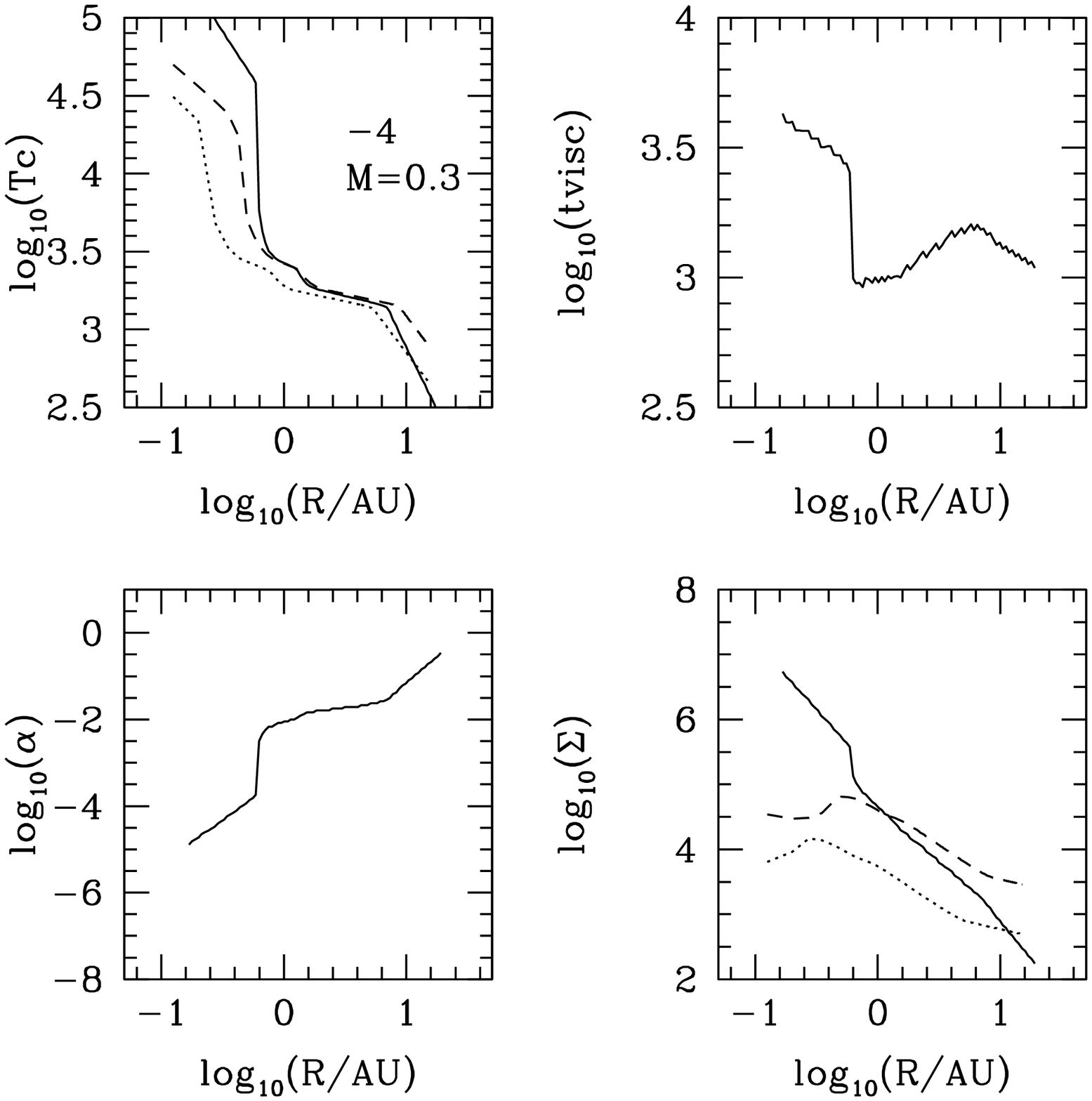} \hfil
\includegraphics[width=0.42\textwidth]{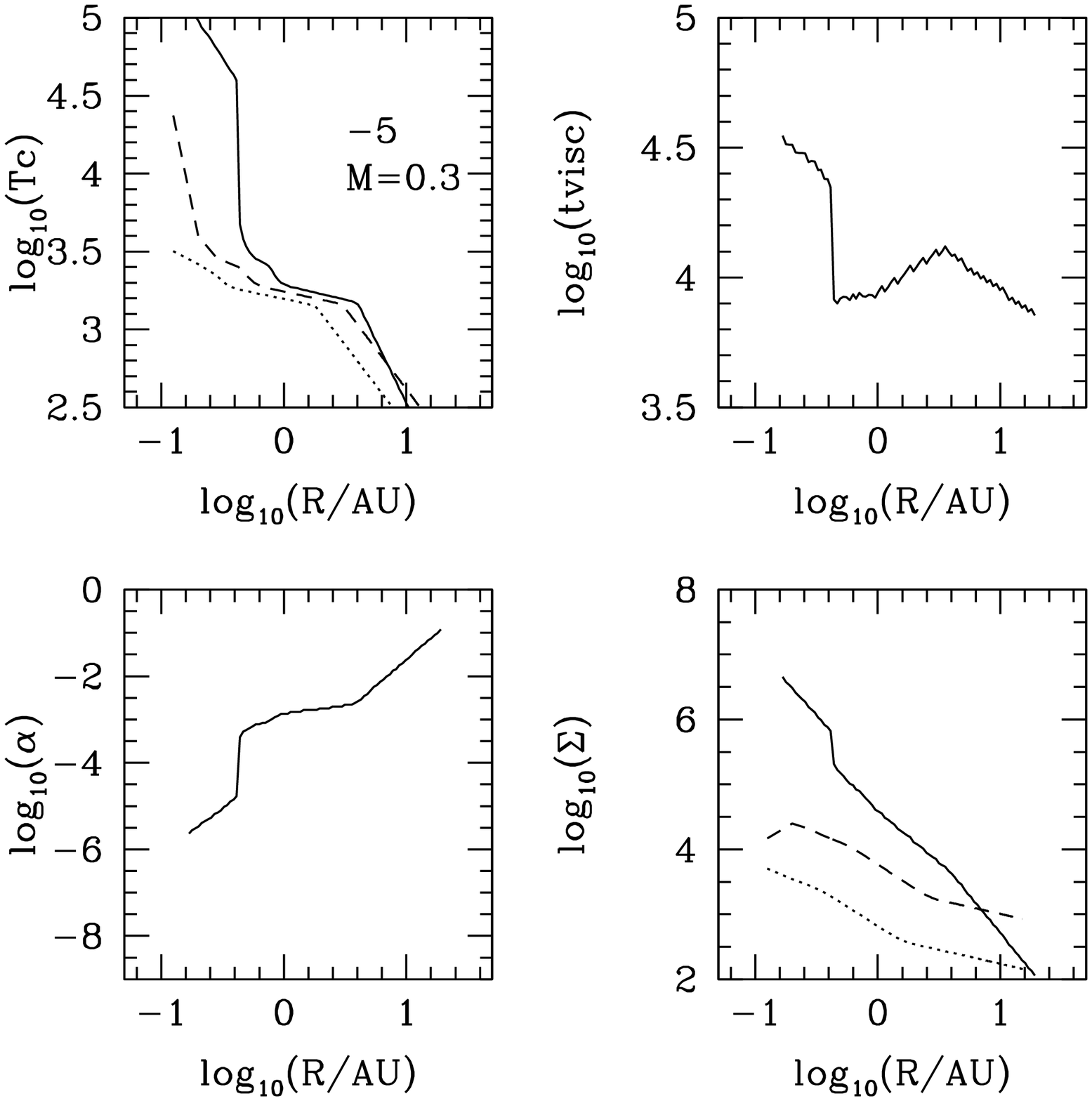} \\
\vskip 0.5cm
\includegraphics[width=0.42\textwidth]{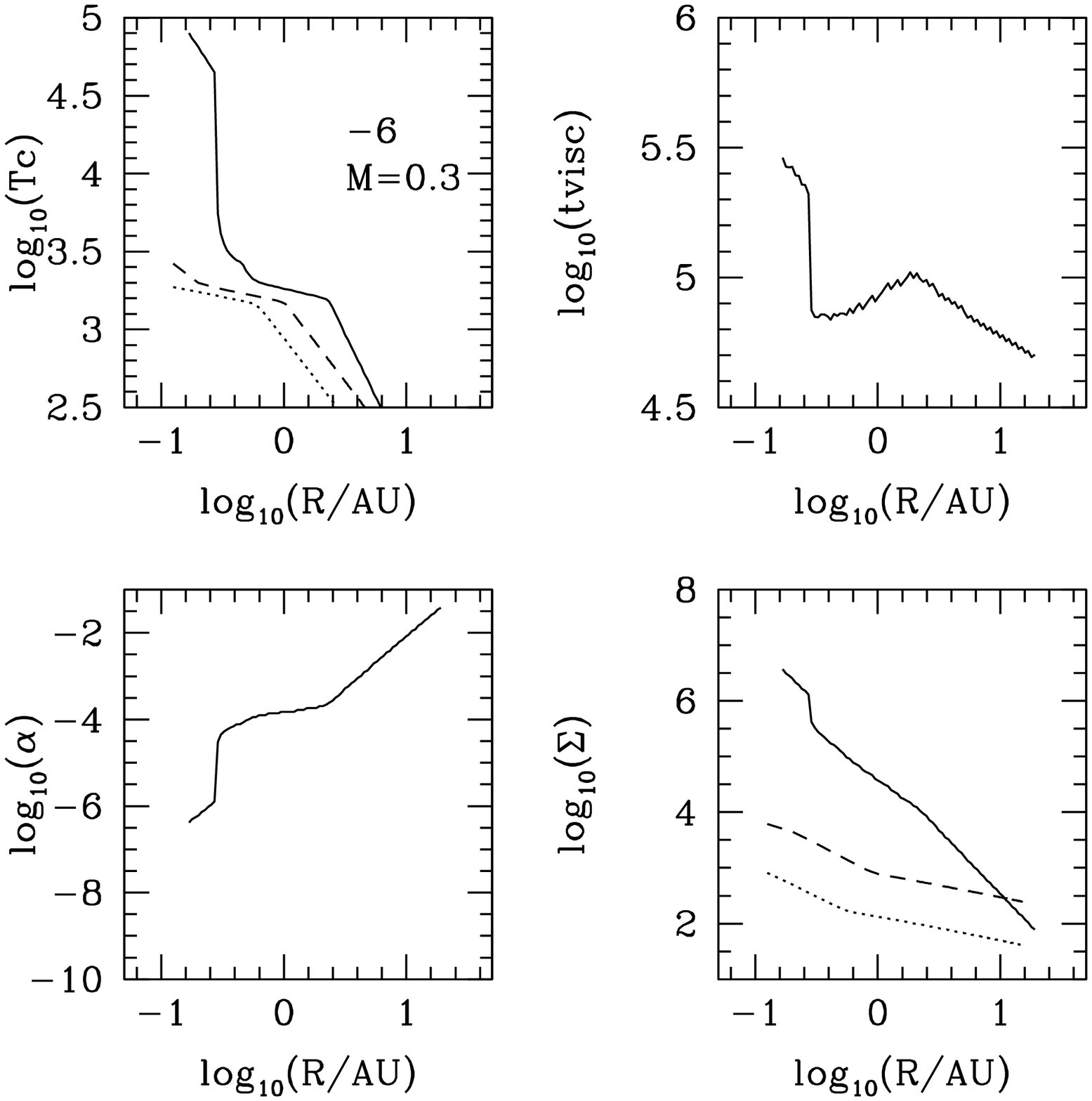} \hfil
\includegraphics[width=0.42\textwidth]{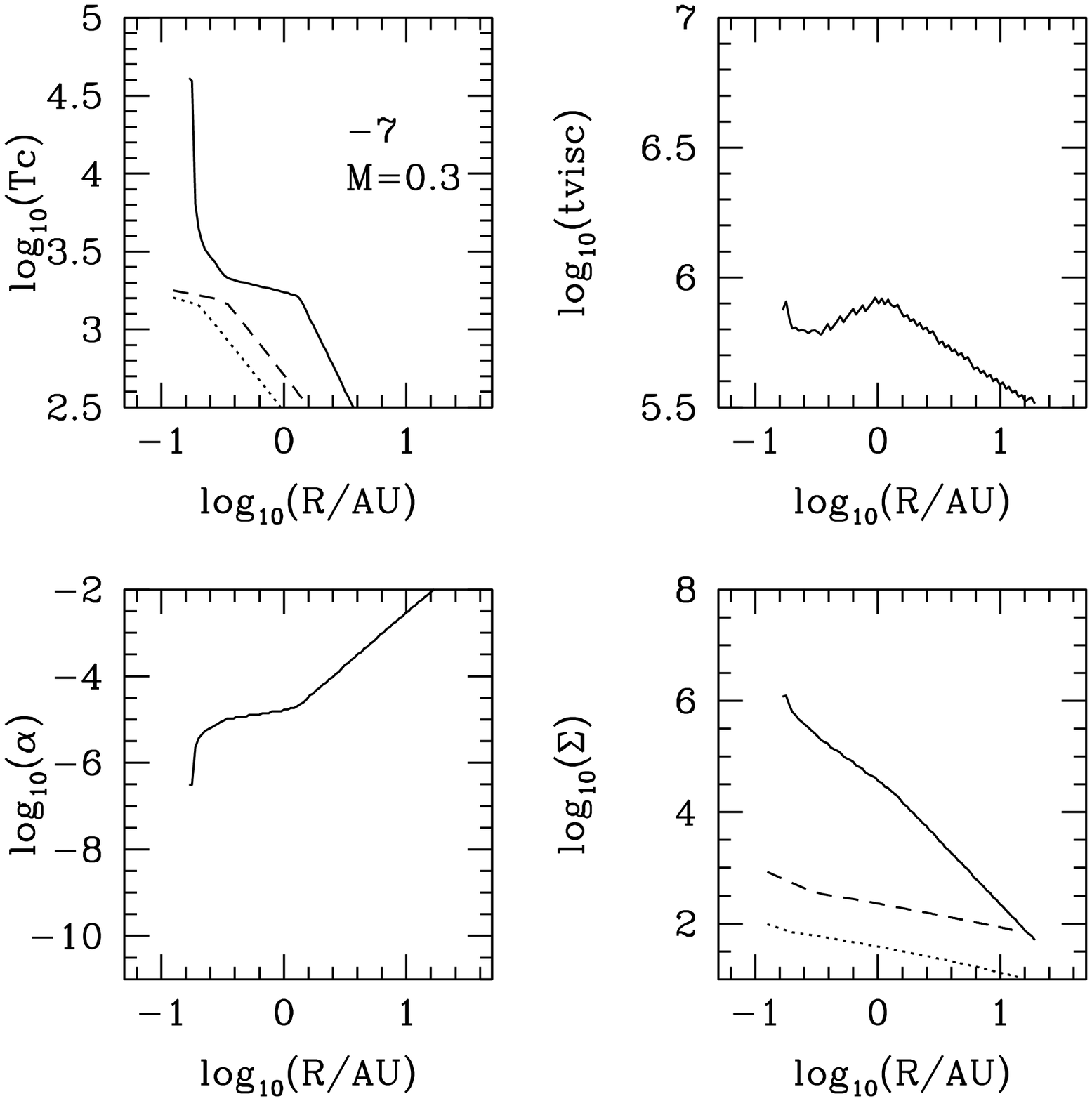} \\
\caption{Same as in Figure \ref{fig:f1} but for a central star mass
of $0.3 \msun$.}
\label{fig:f3}
\end{figure}

\begin{figure}
\includegraphics[width=0.42\textwidth]{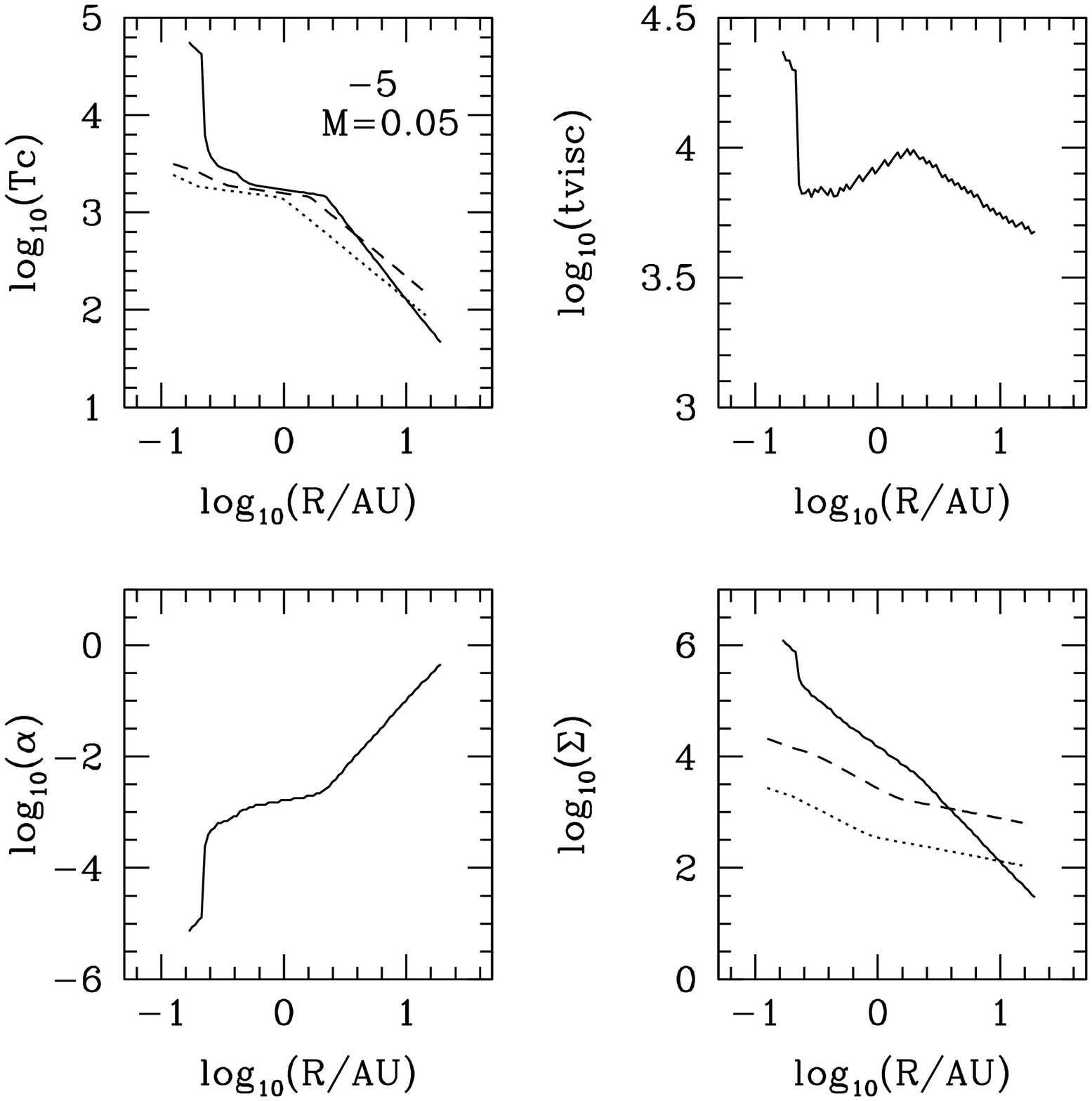} \hfil
\includegraphics[width=0.42\textwidth]{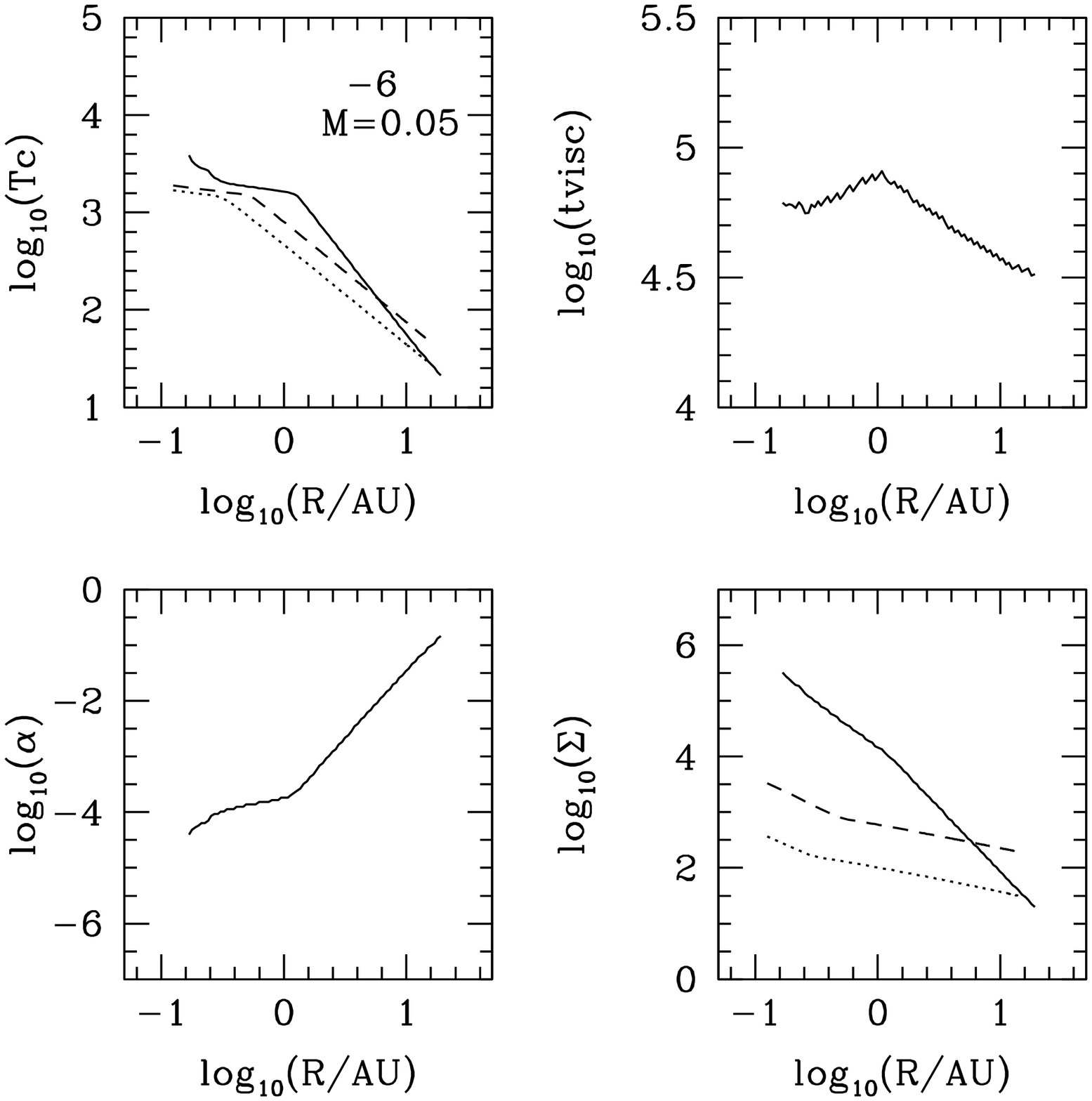} \\
\vskip 0.5cm
\includegraphics[width=0.42\textwidth]{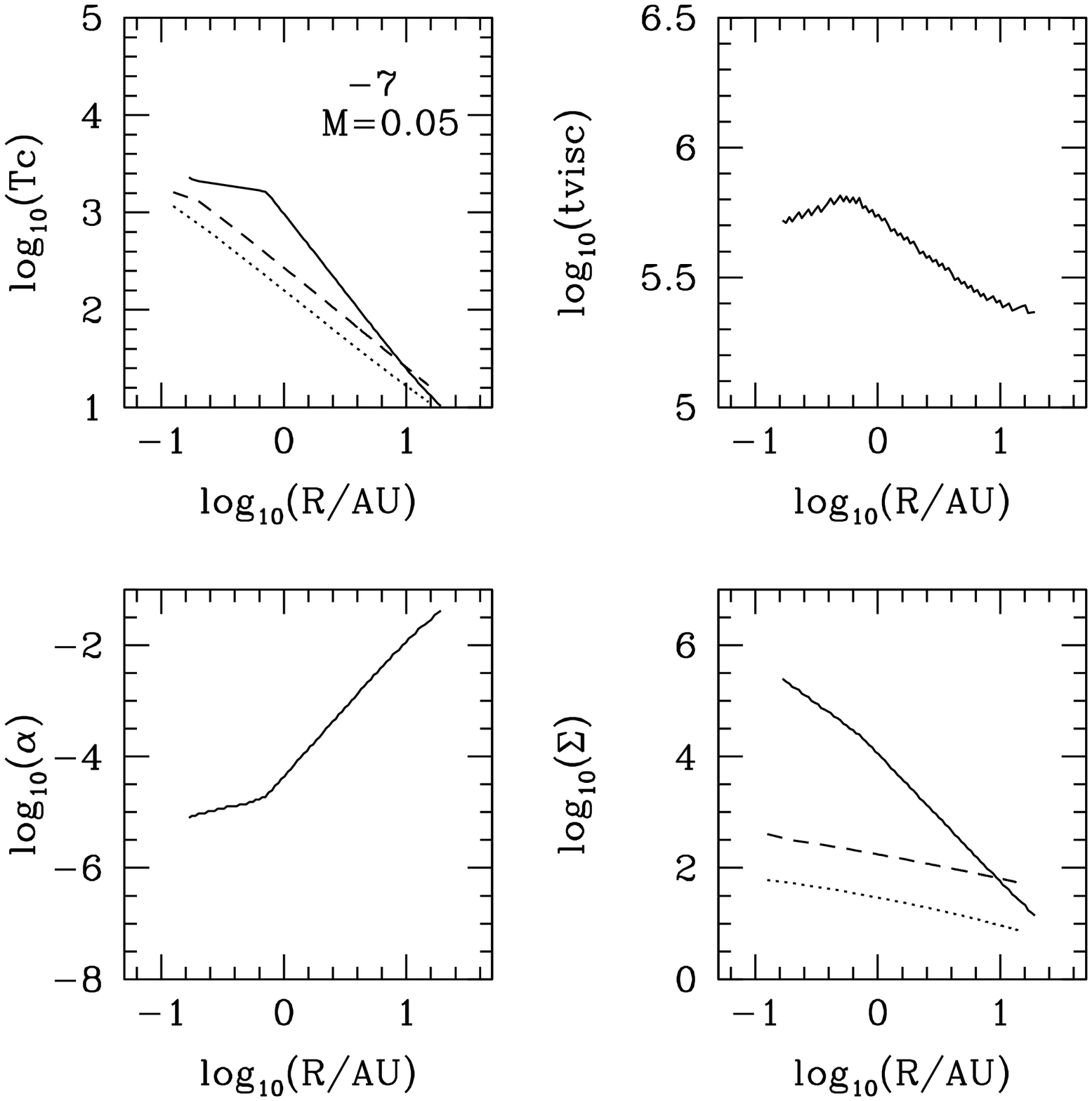} \\
\caption{Same as in Figure \ref{fig:f1} but for a central star mass
of $0.05 \msun$.}
\label{fig:f4}
\end{figure}

\begin{figure}
\includegraphics[width=0.42\textwidth]{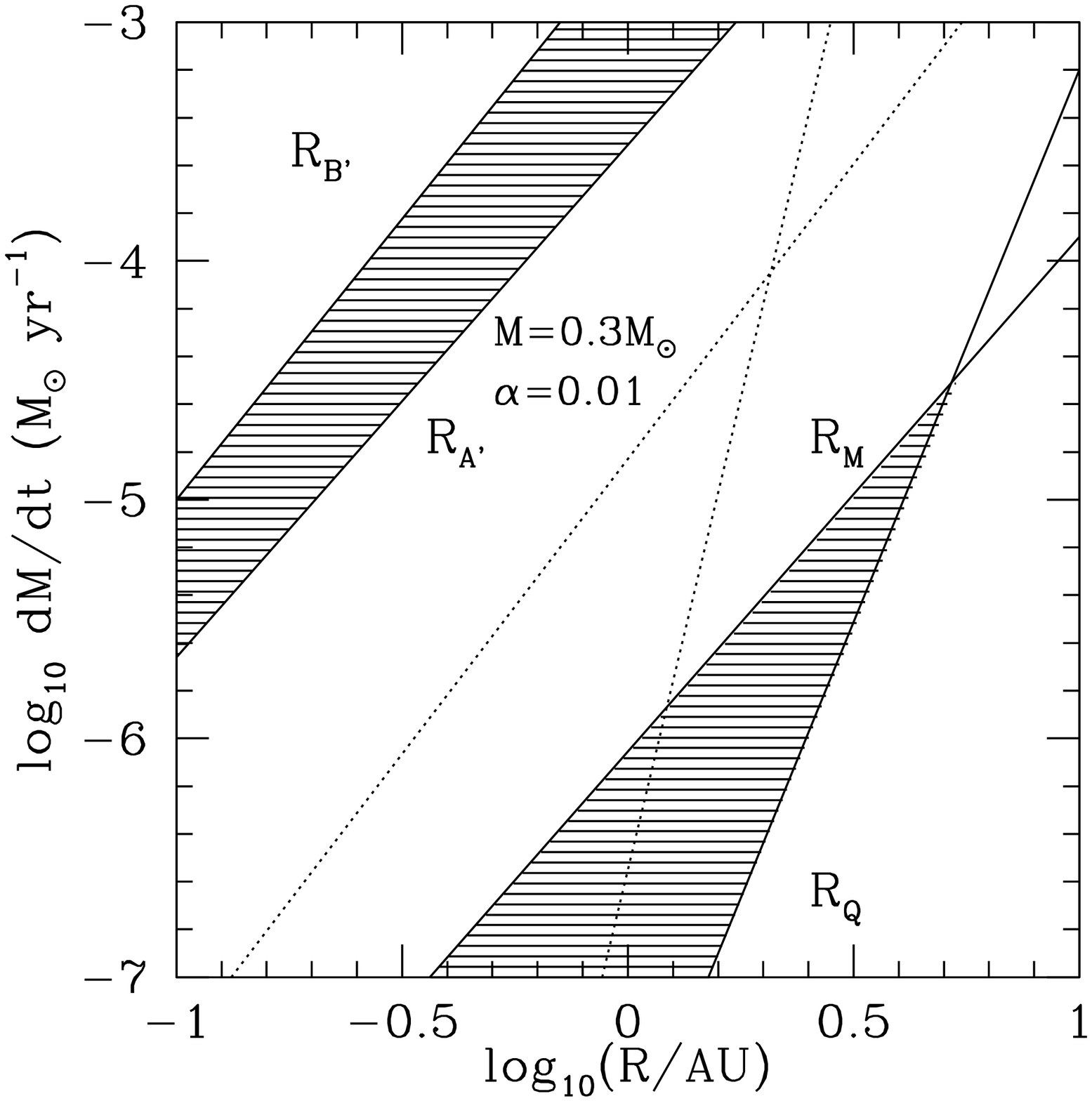} \hfil
\includegraphics[width=0.42\textwidth]{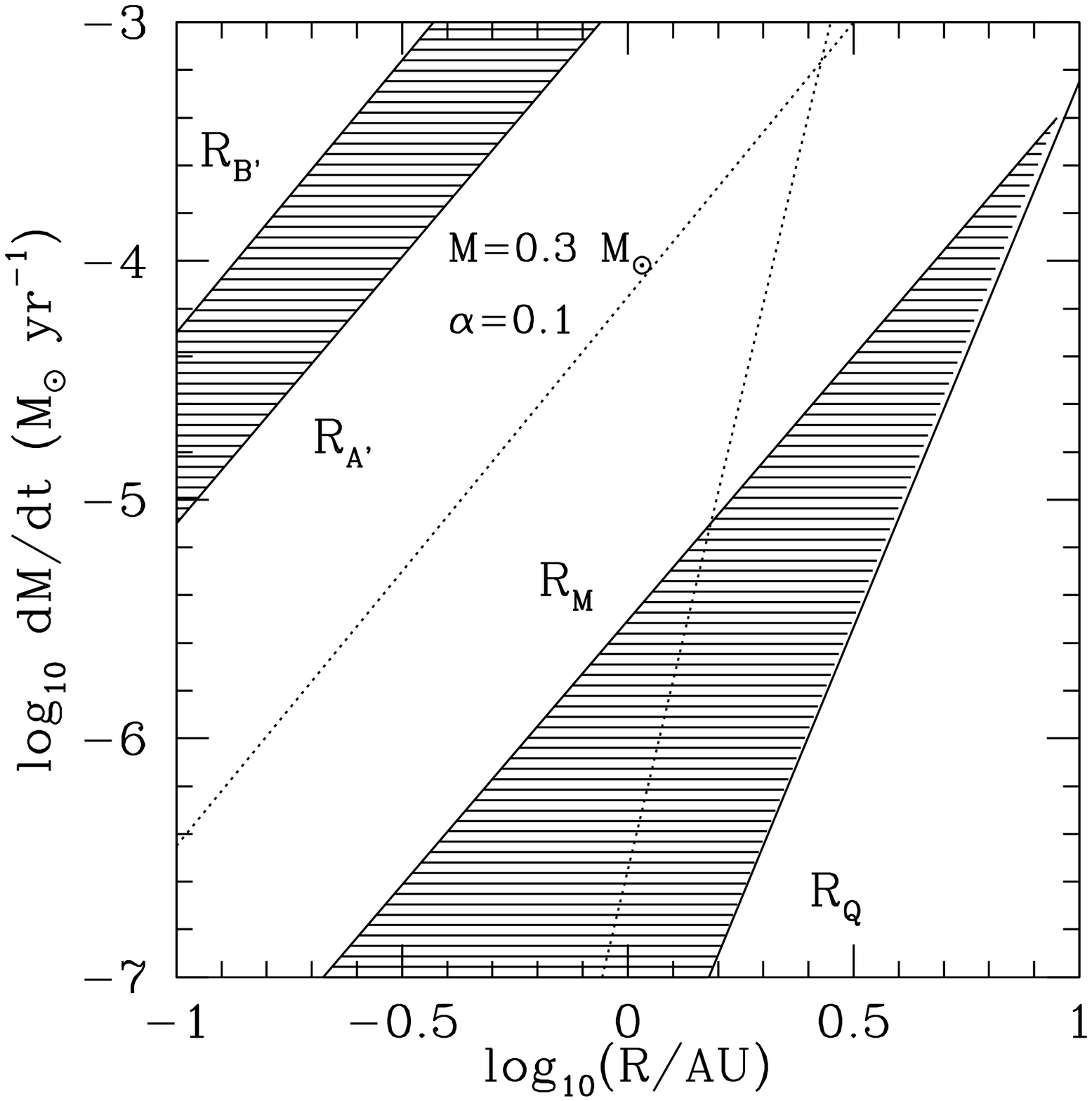} \\
\caption{Same as figure \ref{fig:f2} for
$0.3 \msun$ central star.}
\label{fig:f5}
\end{figure}

\begin{figure}
\includegraphics[width=0.42\textwidth]{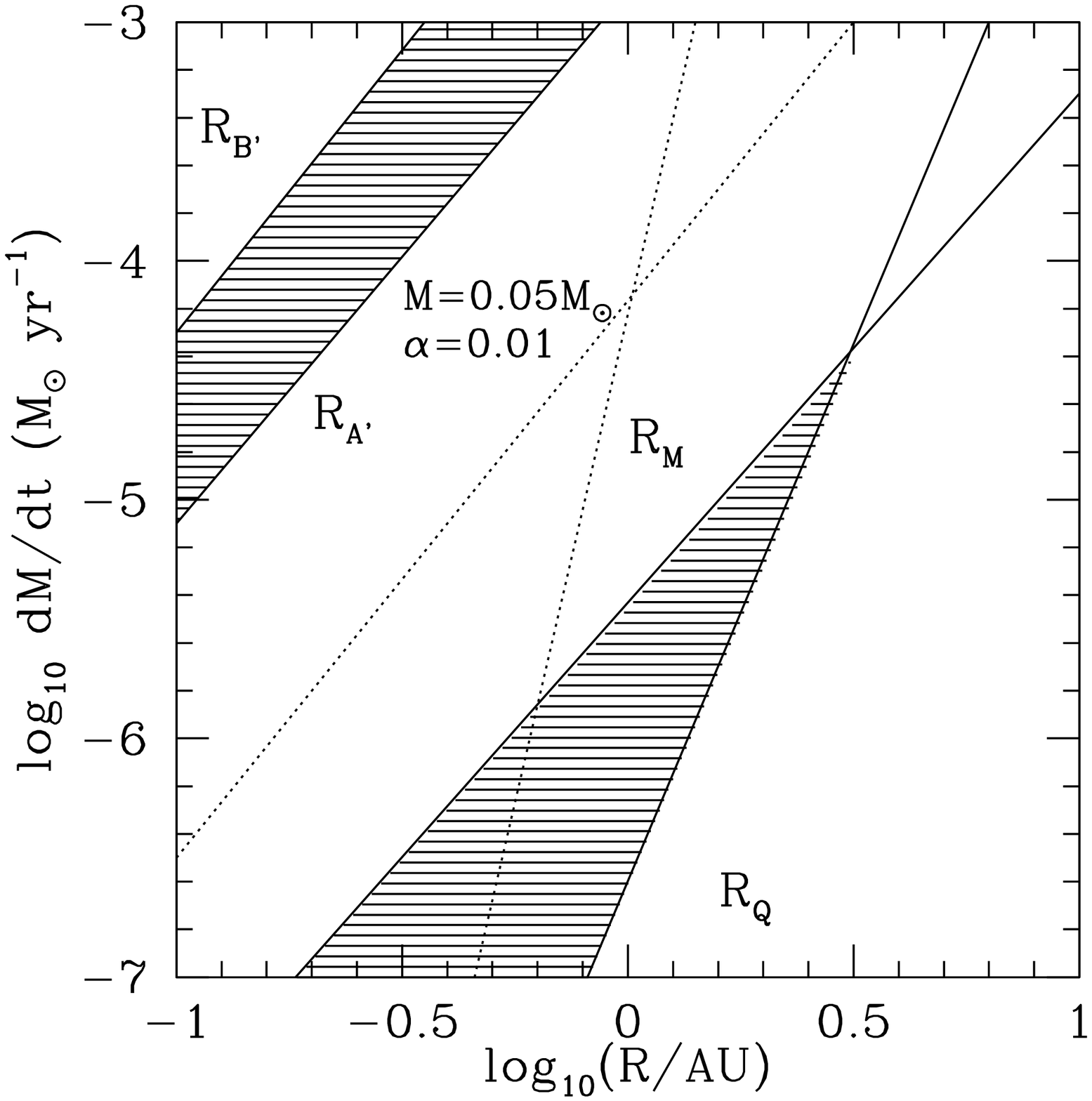} \hfil
\includegraphics[width=0.42\textwidth]{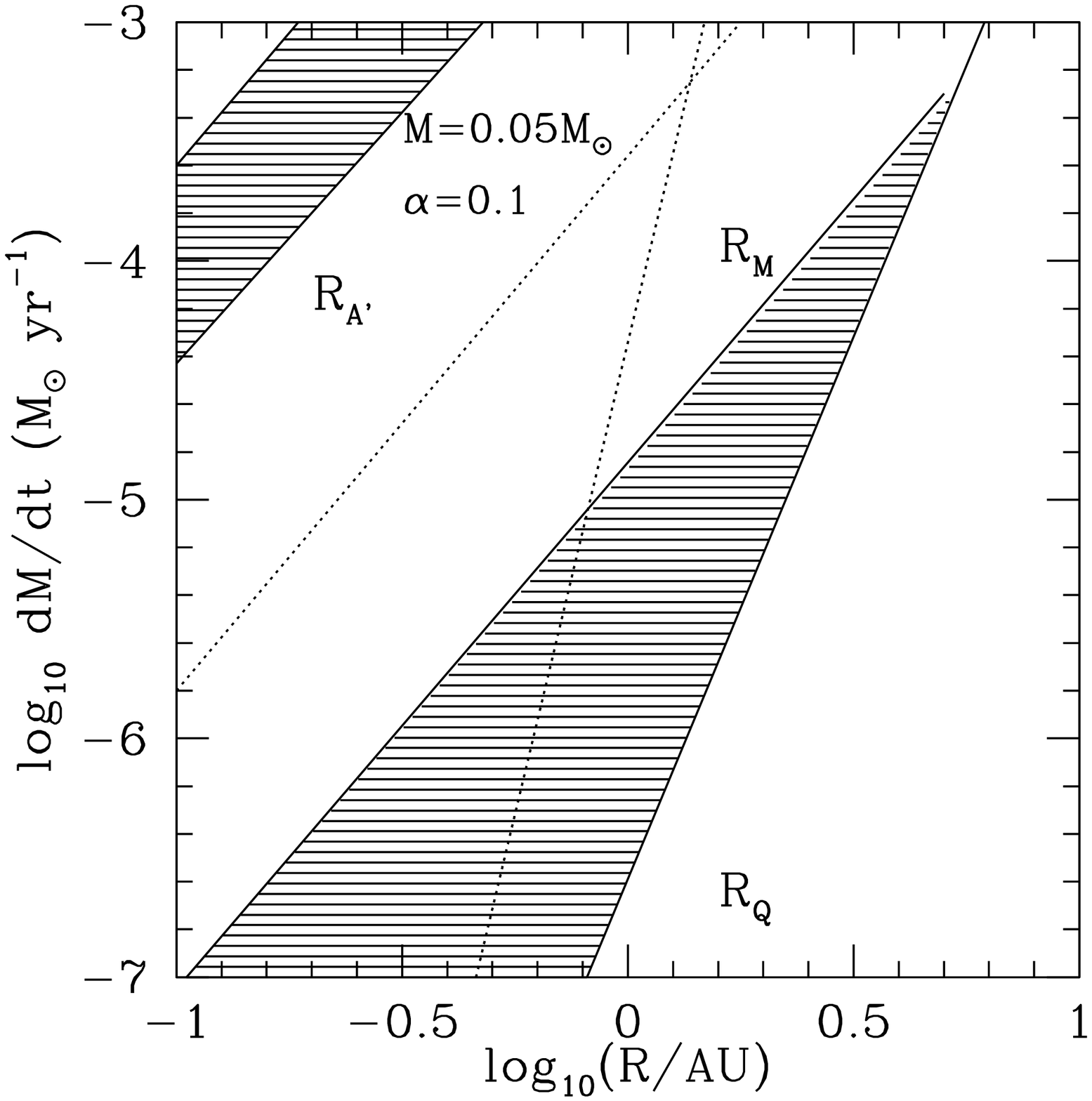} \\
\caption{Same as Figure \ref{fig:f2} for the $0.05 \msun$ central star.}
\label{fig:f6}
\end{figure}

\begin{figure}
\includegraphics[width=0.7\textwidth]{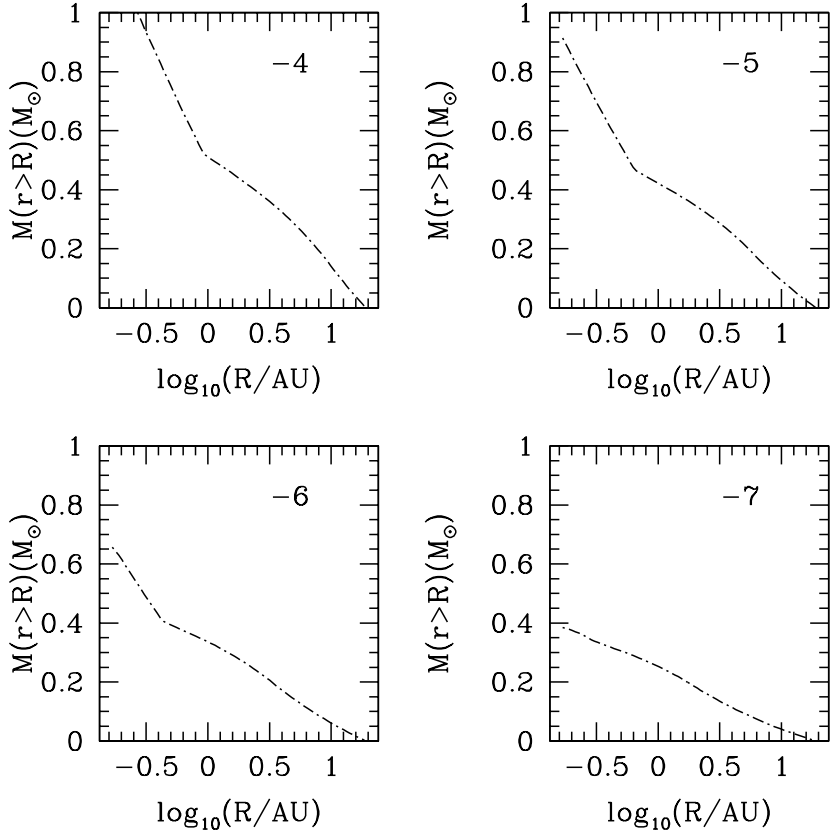}
\caption{The mass of the disk integrated between radius R and the
outer radius of 20 AU for steady-state Q=2 disks, at four accretion
rates - $10^{-4}$, $10^{-5}$, $10^{-6}$, and $10^{-7} \msunyr$. The
 central star mass is $1 \msun$. }
\end{figure}

\begin{figure}
\includegraphics[width=0.7\textwidth]{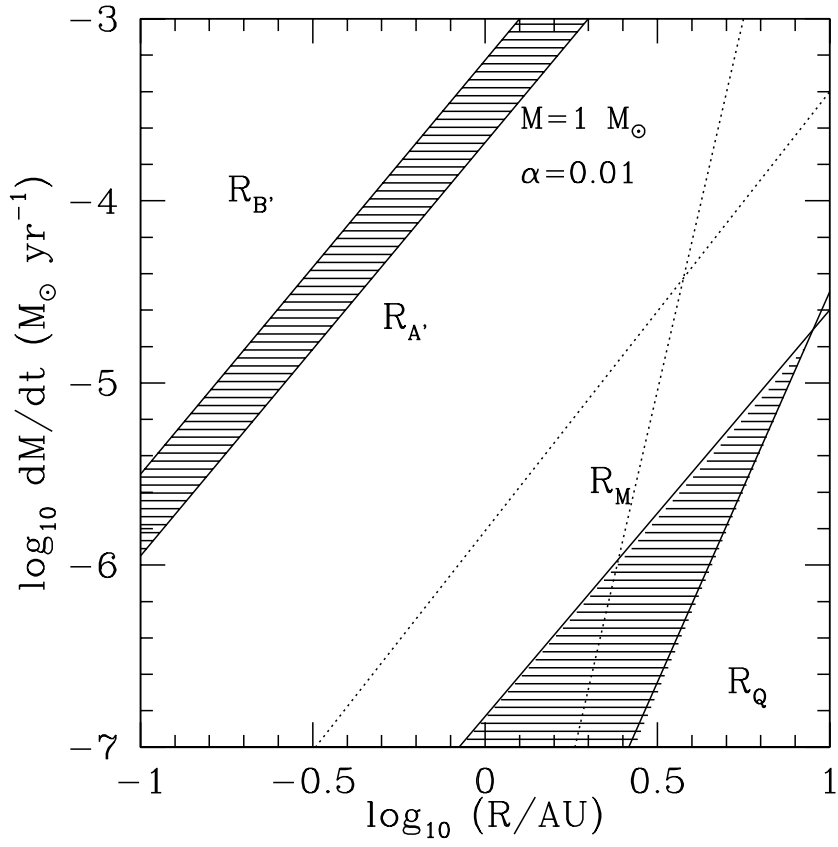}
\caption{Same as Figure \ref{fig:f2} for the parameters of
Armitage et al. (2001) (see text)}
\end{figure}

\begin{figure}
\epsscale{0.5} \plotone{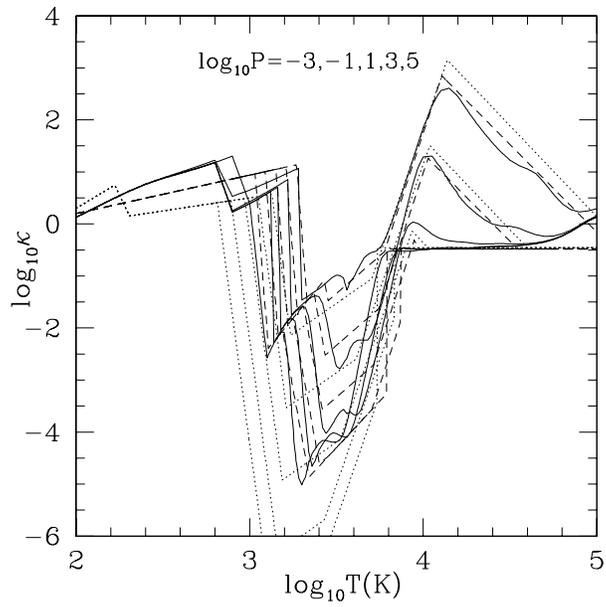} \caption{Rosseland mean opacities:
the dotted lines show the Bell \& Lin (1994) fit, the solid curves show
the
detailed opacity calculation of Zhu \etal (2007, 2008) (solid line), and
the dashed lines show the simple fit to
Zhu \etal opacities (Table 1).} \label{fig:opa}
\end{figure}

\FloatBarrier

\begin{table}
\begin{center}
\caption{Fit to Zhu \etal (2007, 2008) opacity  \label{fitopa}}
\begin{tabular}{clcl}
\tableline\tableline
$\log_{10}T$& $\log_{10}\kappa$ & comments \\
\tableline
$< 0.03 \log_{10}P + 3.12$ & $0.738 \log_{10}T - 1.277$
& grain opacity \\
$< 0.0281  \log_{10}P + 3.19 $&$ -42.98 \log_{10}T + 1.312 \log_{10}P +
135.1$ & grain evaporation \\
$< 0.03 \log_{10}P + 3.28$ & $4.063 \log_{10}T -15.013$ & water vapor
\\
$< 0.00832 \log_{10}P + 3.41$ &$ -18.48 \log_{10}T + 0.676 \log_{10}P
+58.93$ &  \\
$< 0.015 \log_{10}P + 3.7$ & $2.905 \log_{10}T + 0.498 \log_{10}P
-13.995$ & molecular opacities \\
$< 0.04 \log_{10}P + 3.91 $&$ 10.19 \log_{10}T + 0.382\log_{10}P
-40.936$ & H scattering \\
$< 0.28 \log_{10}P + 3.69$ & $-3.36 \log_{10}T + 0.928 \log_{10}P
+12.026$ & bound-free,free-free \\
else \tablenotemark{a} & $-0.48$ & electron scattering \\
\tableline
\end{tabular}
\tablenotetext{a}{with two additional condition to set the boundary:
if $\log_{10}\kappa <$3.586 $\log_{10}T$ -16.85 and $\log_{10}T$ $<$
4, $\log_{10}\kappa=$ 3.586 $\log_{10}T$-16.85; if $\log_{10}T$ $<$
2.9, $\log_{10}\kappa$=0.738 $\log_{10}T$ -1.277}
\end{center}
\end{table}


\begin{thebibliography}

\bibitem[Alexander \& Ferguson(1994)]{alexander94} Alexander, D.~R., \&
Ferguson, J.~W.\ 1994, \apj, 437, 879

\bibitem[Andre et al.(2000)]{2000prpl.conf...59A} Andre, P.,
Ward-Thompson, D., \& Barsony, M.\ 2000, Protostars and Planets IV, 59

\bibitem[Armitage et al.(2001)]{Armitage2001} Armitage, P.~J., Livio,
M., \& Pringle, J.~E.\ 2001, \mnras, 324, 705

\bibitem[Balbus \& Papaloizou(1999)]{1999ApJ...521..650B} Balbus, S.~A.,
\& Papaloizou, J.~C.~B.\ 1999, \apj, 521, 650

\bibitem[Bally et al.(2007)]{2007prpl.conf..215B} Bally, J., Reipurth,
B., \& Davis, C.~J.\ 2007, Protostars and Planets V, 215

\bibitem[Bate et al.(2003)]{2003MNRAS.339..577B} Bate, M.~R., Bonnell,
I.~A., \& Bromm, V.\ 2003, \mnras, 339, 577

\bibitem[Beckwith et al.(2008)]{bhk08} Beckwith, K., Hawley, J.~F., \&
Krolik, J.~H.\ 2008, \apj, 678, 1180

\bibitem[Bell \& Lin(1994)]{1994ApJ...427..987B} Bell, K.~R., \& Lin,
D.~N.~C.\ 1994, \apj, 427, 987

\bibitem[Boley et al.(2006)]{Boley2006} Boley, A.~C.,
Mej{\'{\i}}a, A.~C., Durisen, R.~H., Cai, K., Pickett, M.~K., \&
D'Alessio, P.\ 2006, \apj, 651, 517

\bibitem[Book \& Hartmann(2005)]{2005AAS...207.7417B}
Book, L.~G., \& Hartmann, L.\ 2005, BAAS, 37, 1287

\bibitem[Brandenburg et al.(1995)] {bran95} Brandenburg, A., Nordlund,
A., Stein, R., \& Torkelsson, U.\ 1995, \apj, 446, 741

\bibitem[Cai et al.(2008)]{cai2008} Cai, K., Durisen, R.~H.,
Boley, A.~C., Pickett, M.~K., \& Mej{\'{\i}}a, A.~C.\ 2008, \apj,
673, 1138

\bibitem[Cassen(1993)]{1993LPI....24..261C} Cassen, P.\ 1993, Lunar and
Planetary Institute Conference Abstracts, 24, 261

\bibitem[Castelli(2005)]{castelli05} Castelli, F.\ 2005, Memorie
della Societa Astronomica Italiana Supplement, 8, 34

\bibitem[Castelli \& Kurucz(2004)]{kurucz04} Castelli, F., \& Kurucz,
R.~L.\ 2004, ArXiv Astrophysics e-prints, arXiv:astro-ph/0405087

\bibitem[Clarke \& Syer(1996)]{1996MNRAS.278L..23C} Clarke, C.~J., \&
Syer, D.\ 1996, \mnras, 278, L23

\bibitem[D'Alessio et al.(2001)]{dalessio2001} D'Alessio, P.,
Calvet, N., \& Hartmann, L.\ 2001, \apj, 553, 321

\bibitem[D'Alessio et al.(1998)]{dalessio1998} D'Alessio, P., Canto, J.,
Calvet, N., \& Lizano, S.\ 1998, \apj, 500, 411

\bibitem[Desch(1998)]{1998PhDT........11D} Desch, S.~J.\ 1998,
Ph.D.~Thesis, U. Illinois.

\bibitem[Faulkner et al.(1983)]{Lin83I} Faulkner, J., Lin,
D.~N.~C., \& Papaloizou, J.\ 1983, \mnras, 205, 359

\bibitem[Ferguson et al.(2005)]{ferguson05} Ferguson, J.~W., Alexander,
D.~R., Allard, F., Barman, T., Bodnarik, J.~G., Hauschildt, P.~H.,
Heffner-Wong, A., \& Tamanai, A.\ 2005, \apj, 623, 585

\bibitem[Fleming
\& Stone(2003)]{fleming2003} Fleming, T., \& Stone, J.~M.\ 2003,
\apj, 585, 908

\bibitem[Foster \& Chevalier(1993)]{1993ApJ...416..303F} Foster, P.~N.,
\& Chevalier, R.~A.\ 1993, \apj, 416, 303

\bibitem[Fromang \& Papaloizou(2007)]{fp07a} Fromang, S., \& Papaloizou,
J.\ 2007 \aa, 476, 1113 (FP07)

\bibitem[Fromang et al.(2007)]{fp07b} Fromang, S., Papaloizou, J.,
Lesur, G., \& Heinemann, T.\ 2007, \aap, 476

\bibitem[Furlan et al.(2008)]{2008ApJS..176..184F} Furlan, E., et al.\
2008, \apjs, 176, 184

\bibitem[Gammie(1996)]{1996ApJ...457..355G} Gammie, C.~F.\ 1996, \apj,
457,
355

\bibitem[Gammie(2001)]{2001ApJ...553..174G} Gammie, C.~F.\ 2001, \apj,
553, 174

\bibitem[Guan(2008)]{2008apjinpress} Guan,  Gammie, C.~F. \etal 2008, in
preparation

\bibitem[Johnson \& Gammie(2003)]{2003ApJ...597..131J} Johnson, B.~M.,
\& Gammie, C.~F.\ 2003, \apj, 597, 131

\bibitem[Gammie(1999)]{1999ASPC..160..122G} Gammie, C.~F.\ 1999,
ASP Conference series 160, 122

\bibitem[Guan et al.(2008)]{ggsj} Guan, X., Gammie, C.~F., Simon, J.,
and Johnson, B.~M. 2008, \apj, in prep.

\bibitem[Glassgold et al.(1997)]{1997ApJ...480..344G} Glassgold, A.~E.,
Najita, J., \& Igea, J.\ 1997, \apj, 480, 344

\bibitem[Hartmann et al.(1994)]{1994ApJ...430L..49H} Hartmann, L., Boss,
A., Calvet, N., \& Whitney, B.\ 1994, \apjl, 430, L49

\bibitem[Hartmann et al.(1997)]{1997ApJ...475..770H} Hartmann, L.,
Cassen,
P., \& Kenyon, S.~J.\ 1997, \apj, 475, 770

\bibitem[Hartmann \& Kenyon(1996)]{1996ARA&A..34..207H} Hartmann, L., \&
Kenyon, S.~J.\ 1996, \araa, 34, 207

\bibitem[Hawley et al.(1995)]{hgb95} Hawley, J.~F., Gammie, C.~F., \&
Balbus, S.~A.\ 1995, \apj, 440, 742

\bibitem[Hawley et al.(1996)]{hgb96} Hawley, J.~F., Gammie, C.~F., \&
Balbus, S.~A.\ 1996, \apj, 464, 690

\bibitem[Henriksen et al.(1997)]{1997A&A...323..549H} Henriksen, R.,
Andre, P., \& Bontemps, S.\ 1997, \aap, 323, 549

\bibitem[Herbig(1977)]{1977ApJ...217..693H} Herbig, G.~H.\ 1977, \apj,
217,
693

\bibitem[Hirose et al.(2004)]{hkvh05} Hirose, S., Krolik, J., De
Villiers, J.-P., \& Hawley, J.\ 2005, \apj, 606, 1083

\bibitem[Ilgner
\& Nelson(2006a)]{2006A&A...445..205I} Ilgner, M., \& Nelson, R.~P.\ 2006a, \aap, 445, 205

\bibitem[Ilgner
\& Nelson(2006b)]{2006A&A...445..223I} Ilgner, M., \& Nelson, R.~P.\ 2006b, \aap, 445, 223

\bibitem[Ilgner
\& Nelson(2006c)]{2006A&A...455..731I} Ilgner, M., \& Nelson, R.~P.\ 2006c, \aap, 455, 731

\bibitem[Ilgner
\& Nelson(2008)]{2008A&A...483..815I} Ilgner, M., \& Nelson, R.~P.\ 2008, \aap, 483, 815

\bibitem[Johansen
\& Levin(2008)]{2008A&A...490..501J} Johansen, A., \& Levin, Y.\ 2008, \aap, 490, 501

\bibitem[Kenyon et al.(1993)]{1993ApJ...414..676K} Kenyon, S.~J.,
Calvet,
N., \& Hartmann, L.\ 1993, \apj, 414, 676


\bibitem[Kenyon et al.(1990)]{1990AJ.....99..869K} Kenyon, S.~J.,
Hartmann,
L.~W., Strom, K.~M., \& Strom, S.~E.\ 1990, \aj, 99, 869

\bibitem[Kenyon et al.(1994)]{1994AJ....108..251K} Kenyon, S.~J., Gomez,
M., Marzke, R.~O., \& Hartmann, L.\ 1994, \aj, 108, 251

\bibitem[King et al.(2007)]{2007MNRAS.376.1740K} King, A.~R., Pringle,
J.~E., \& Livio, M.\ 2007, \mnras, 376, 1740

\bibitem[Kurucz et al.(1974)]{kurucz74} Kurucz, R.~L., Peytremann, E.,
\& Avrett, E.~H.\ 1974, Washington : Smithsonian Institution : for sale
by the Supt.~of Docs., U.S.~Govt.~Print.~Off., 1974., 37

\bibitem[Lesur \& Longaretti(2007)]{ll07} Lesur, G., \& Longaretti,
P.-Y.\ 2007, \mnras, 378, 1471

\bibitem[Lin et al.(1985)]{Lin85III} Lin, D.~N.~C., Faulkner, J., \&
Papaloizou, J.\ 1985, \mnras, 212, 105

\bibitem[Lin
\& Papaloizou(1980)]{linpapa1980} Lin, D.~N.~C., \& Papaloizou, J.\
1980, \mnras, 191, 37

\bibitem[Lin \& Papaloizou(1985)]{1985prpl.conf..981L} Lin, D.~N.~C., \&
Papaloizou, J.\ 1985, Protostars and Planets II, 981

\bibitem[Lodato \& Clarke(2004)]{2004MNRAS.353..841L} Lodato, G., \&
Clarke, C.~J.\ 2004, \mnras, 353, 841

\bibitem[McKinney \& Narayan(2007)]{mn07} McKinney, J.~C., \& Narayan,
R.\ 2007, \mnras, 375, 513

\bibitem[Muzerolle et al.(1998)]{1998AJ....116.2965M} Muzerolle, J.,
Hartmann, L., \& Calvet, N.\ 1998, \aj, 116, 2965

\bibitem[Myers et al.(1998)]{1998ApJ...492..703M} Myers, P.~C., Adams,
F.~C., Chen, H., \& Schaff, E.\ 1998, \apj, 492, 703

\bibitem[Tannirkulam et al.(2008)]{2008arXiv0808.1728T} Tannirkulam,
A., et
al.\ 2008, ArXiv e-prints, 808, arXiv:0808.1728

\bibitem[Sano et al.(2000)]{2000ApJ...543..486S} Sano, T., Miyama,
S.~M.,
Umebayashi, T., \& Nakano, T.\ 2000, \apj, 543, 486

\bibitem[Sbordone et al.(2004)]{kurucz042} Sbordone, L., Bonifacio, P.,
Castelli, F., \& Kurucz, R.~L.\ 2004, Memorie della Societa Astronomica
Italiana Supplement, 5, 93

\bibitem[Shu et al.(1987)]{1987ARA&A..25...23S} Shu, F.~H., Adams,
F.~C., \& Lizano, S.\ 1987, \araa, 25, 23

\bibitem[Stahler(1988)]{1988ApJ...332..804S} Stahler, S.~W.\ 1988, \apj,
332, 804

\bibitem[Terquem(2008)]{terquem2008} Terquem, C.~E.~J.~M.~L.~J.\
2008, arXiv:0808.3897

\bibitem[Turner
\& Sano(2008)]{turner2008} Turner, N.~J., \& Sano, T.\ 2008, \apjl,
679, L131

\bibitem[Vallee(2004)]{va04}Vallee, J.~P.\ 2004, NewA Rev., 48, 763

\bibitem[van Ballegooijen(1989)]{vanball} van Ballegooijen, A.~A.\ 1989,
Accretion Disks and Magnetic Fields in Astrophysics, 156, 99

\bibitem[Vorobyov \& Basu(2005)]{2005ApJ...633L.137V} Vorobyov, E.~I.,
\& Basu, S.\ 2005, \apjl, 633, L137

\bibitem[Vorobyov \& Basu(2006)]{2006ApJ...650..956V} Vorobyov, E.~I.,
\& Basu, S.\ 2006, \apj, 650, 956

\bibitem[Vorobyov \& Basu(2008)]{2008VB} Vorobyov, E.~I., \& Basu, S.
arXiv:0802.2242v1

bibitem[White et al.(2007)]{2007prpl.conf..117W} White, R.~J., Greene,
T.~P., Doppmann, G.~W., Covey, K.~R.,
\& Hillenbrand, L.~A.\ 2007, Protostars and Planets V, 117

\bibitem[White \& Hillenbrand(2004)]{2004ApJ...616..998W}
White, R.~J., \& Hillenbrand, L.~A.\ 2004, \apj, 616, 998

\bibitem[Zhu et al.(2007)]{zhu2007} Zhu, Z., Hartmann, L., Calvet, N.,
Hernandez, J., Muzerolle, J., \& Tannirkulam, A.-K.\ 2007, \apj, 669,
483

\bibitem[Zhu et al.(2008)]{2008arXiv0806.3715Z} Zhu, Z., Hartmann, L.,
Calvet, N., Hernandez, J., Tannirkulam, A.-K.,
\& D'Alessio, P.\ 2008, ArXiv e-prints, 806, arXiv:0806.3715
(ApJ, in press)

\end{thebibliography}
\end{document}